\documentclass{aa}
\usepackage{epsfig}
\usepackage{natbib}
\begin{document}
\newcommand{\bea}{\begin{eqnarray}}    
\newcommand{\eea}{\end{eqnarray}}      
\newcommand{\be}{\begin{equation}}
\newcommand{\ee}{\end{equation}}
\newcommand{\bef}{\begin{figue}}
\newcommand{\eef}{\end{figure}}
\newcommand{\etal}{et al.}
\newcommand{\kms}{\,{\rm km}\;{\rm s}^{-1}}
\newcommand{\hubunits}{\,\kms\;{\rm Mpc}^{-1}}
\newcommand{\hmpc}{\,h^{-1}\;{\rm Mpc}}
\newcommand{\hkpc}{\,h^{-1}\;{\rm kpc}}
\newcommand{\msun}{M_\odot}
\newcommand{\K}{\,{\rm K}}
\newcommand{\cm}{{\rm cm}}
\newcommand{\cd}{{\langle n(r) \rangle_p}}
\newcommand{\Mpc}{{\rm Mpc}}
\newcommand{\kpc}{{\rm kpc}}
\newcommand{\xir}{{\xi(r)}}
\newcommand{\xrp}{{\xi(r_p,\pi)}}
\newcommand{\xsirpi}{{\xi(r_p,\pi)}}
\newcommand{\wrp}{{w_p(r_p)}}
\newcommand{\gr}{{g-r}}
\newcommand{\Navg}{N_{\rm avg}}
\newcommand{\Mmin}{M_{\rm min}}
\newcommand{\fiso}{f_{\rm iso}}
\newcommand{\Mr}{M_r}
\newcommand{\rp}{r_p}
\newcommand{\zmax}{z_{\rm max}}
\newcommand{\zmin}{z_{\rm min}}

\def\eg{{e.g.}}
\def\ie{{i.e.}}
\def\spose#1{\hbox to 0pt{#1\hss}}
\def\ltapprox{\mathrel{\spose{\lower 3pt\hbox{$\mathchar"218$}}
\raise 2.0pt\hbox{$\mathchar"13C$}}}
\def\gtapprox{\mathrel{\spose{\lower 3pt\hbox{$\mathchar"218$}}
\raise 2.0pt\hbox{$\mathchar"13E$}}}
\def\inapprox{\mathrel{\spose{\lower 3pt\hbox{$\mathchar"218$}}
\raise 2.0pt\hbox{$\mathchar"232$}}}

\title{Large-scale fluctuations in the distribution of galaxies
from the Two Degree Field Galaxy Redshift Survey}

\subtitle{}

\author{Francesco Sylos Labini \inst{1,2}, Nikolay L. Vasilyev \inst{3}, Yurij 
V. Baryshev \inst{3}}

\titlerunning{Galaxy fluctuations from the 2dFGRS}
\authorrunning{Sylos Labini, Vasilyev, Baryshev}

\institute{ 
Centro Studi e Ricerche Enrico Fermi, Via Panisperna 89 A, 
Compendio del Viminale, 00184 Rome, Italy
\and Istituto dei Sistemi Complessi CNR, 
Via dei Taurini 19, 00185 Rome, Italy.
\and 
Institute of Astronomy, St.Petersburg 
State University, Staryj Peterhoff, 198504,
St.Petersburg, Russia
}

\date{Received / Accepted}

\abstract{ We study statistical properties of galaxy structures in
  several samples extracted from the 2dF Galaxy Redshift Survey.  In
  particular, we measured conditional fluctuations by means of the
  scale-length method and determined their probability
  distribution. In this way we find that galaxy distribution in these
  samples is characterized by large amplitude fluctuations with a
  large spatial extension, whose size is only limited by the sample's
  boundaries.  These fluctuations are quite typical and persistent in
  the sample's volumes, and they are detected in two independent
  regions in the northern and southern galactic caps.  We discuss the
  relation of the scale-length method to several statistical
  quantities, such as counts of galaxies as a function of redshift and
  apparent magnitude. We confirm previous results, which have
  determined by magnitude and redshift counts that there are
  fluctuations of about $30\%$ between the southern and the northern
  galactic caps and we relate explicitly these counts to structures in
  redshift space. We show that the estimation of fluctuation amplitude
  normalized to the sample density is biased by systematic effects,
  which we discuss in detail.  We consider the type of fluctuations
  predicted by standard cosmological models of structure formation in
  the linear regime and, to study nonlinear clustering, we analyze
  several samples of mock-galaxy catalogs generated from the
  distribution of dark matter in cosmological N-body simulations.  In
  this way we conclude that the galaxy fluctuations present in these
  samples are too large in amplitude and too extended in space to be
  compatible with the predictions of the standard models of structure
  formation.
 } \maketitle


\section{Introduction}
\label{sec:intro}

In one of his seminal papers \citet{devac1970} put into a historical
perspective the problem of galaxy large-scale structures and the
question about the scale where galaxy distribution turns to
homogeneity\footnote{Note that de Vaucouleurs simply considered the
  scale where galaxies approaches a random distribution.  A
  distribution can be homogeneous, i.e.  with a well-defined mean
  density, without being random; i.e. there can be small amplitude
  fluctuations with weak long-range correlations.}.  He points out
that observations have first found that galaxies are not randomly
distributed, then that in the fifties the same property was assigned
to cluster centers, and finally that at the end of the sixties the
discovery of super-clusters has still enlarged the scale of structures
in the universe, thus pushing the scale where the approach to
homogeneity occurs to larger and larger scales.

In the past twenty years many observations have been dedicated to the
study of the large-scale distributions of galaxies
\citep{cfa1,cfa2,pp,ssrs2,lcrs,sdss,colless01}.  Despite the fact that
large-scale galaxy structures, of about several hundred Mpc/h
size~\footnote{We use $H_0=100 h$ km/sec/Mpc for the value of the
  Hubble constant.}, have been observed
\citep{dlhg85,gh89,pp,sloangreatwall} to be the typical feature of the
distribution of visible matter in the local universe, the statistical
analysis measuring their properties has identified a characteristic
scale that has only slightly changed since its discovery forty years
ago in angular catalogs. This scale, $r_0$, was measured to be the one
at which fluctuations in the galaxy density field are about twice the
value of the sample density and it was indeed determined to be $r_0
\approx$ 5 Mpc/h in the Shane and Wirtanen angular catalog
\citep{tk69}. Subsequent measurements of this scale --- see
e.g. \citet{dp83,davis88,park,benoist,norbergxi01,
  norbergxi02,zehavietal02,zehavietal04} --- found a similar value,
although in several samples larger values of $r_0$ have been found
(i.e., $r_0 \approx 6-12$ Mpc/h).  This variation was then ascribed to
a luminosity dependent effect --- see e.g.
\citet{davis88,park,benoist,zehavietal02}.

 Theoretical models of galaxy formation, like the cold dark matter
 (CDM) one \citep[see][]{peacock} are able to predict the scale $r_0$
 once it is given the amplitude and correlation properties of
 fluctuations of the initial conditions in the early universe.  The
 normalization of the matter initial condition can be obtained by
 measuring the amplitude and correlation properties of the
 anisotropies of the cosmic microwave background radiation
 (CMBR). Then by calculating the evolution of small density
 fluctuations in the linear perturbation analysis of a
 self-gravitating fluid in an expanding universe, it is possible to
 predict the scale $r_0$ today. This turns out, in current models such
 as the CDM ones, to be $r_0 \approx 5$ Mpc/h \citep{springel05}.  On
 scales $r<r_0$ models are unable to make precise predictions of the
 shape of the correlation function because gravitational clustering in
 the nonlinear regime is difficult to be treated. Gravitational N-body
 simulations are then used to investigate structure formation in the
 nonlinear phase.  In addition, given that models predict, for
 $r>r_0$, a precise type of small amplitude fluctuations, it is
 possible to simply relate, by using the linear perturbation analysis
 mentioned above, the properties of fluctuations in the present matter
 density field to those in the initial conditions.  In a certain range
 of scales greater than $r_0$, small amplitude.  fluctuations should
 have still positive correlations. Particularly, for $r_0 < r <r_c$,
 fluctuations have very small amplitude and weak positive correlations
 \cite[see][]{cdm_theo}. On even larger scales $r>r_c$ (where this is
 estimated from CMBR measurements to be $r_c \approx 100$ Mpc/h), all
 models predict that the matter density field presents {\it
   anti-correlations} that tend to zero with a (negative) power-law
 behavior of the type $-r^{-4}$ \citep{glass,cdm_theo}. This negative
 power-law tail corresponds in real-space to the linear dependence of
 the matter power-spectrum (PS) on the wave-number; i.e., $P(k) \sim
 k$.  The former represents a behavior that can be interpreted as a
 consistency requirement for the properties of density fluctuations in
 Friedmann-Robertson-Walker models \citep{glass}.  Because of the
 change in sign of the correlation function at $r_c$, this
 length-scale represents the cut-off in the size of weak amplitude
 structures in standard models. Thus, in the regime where fluctuations
 are small and have weak positive correlations; i.e., for $r_0 < r <
 r_c$, the present matter-density field reflects the imprint of the
 initial conditions.

The fundamental test for current models of galaxy formation then
concerns whether density fluctuations on large scales (i.e., $r>10$
Mpc/h) have small amplitude or not. Another important question
concerns the detection of anti-correlations on scales of $r > r_c
\approx 100$ Mpc/h \citep{cdm_theo}. The primary problem to be
considered in this respect concerns the statistical methods used to
measure the amplitude of fluctuations and the range of
correlations. There has been  intense debate in the past decade
concerning this crucial point
\citep{slmp98,rees,hogg,book,bt05,paper1}. 
 Before one determines the amplitude of fluctuations in a given volume
 with respect to the sample density, one must have firstly tested that
 the former quantity is stable; i.e., that it does not depend on the
 sample size and/or it does not present large fluctuations in
 different samples containing the same type of objects.  Indeed, in
 case the distribution presents structures and fluctuations on all
 scales in a given sample (i.e., it is inhomogeneous) the sample
 density is not a well-defined descriptor \citep{slmp98,paper1}. In
 this situation all statistical quantities that are normalized to the
 sample density are affected by systematic effects. For this reason,
 prior to the characterization of fluctuations with respect to the
 sample density, a fundamental test consists in measuring conditional
 correlation properties \citep{book}. It has been found that
 conditional statistical quantities, such as the conditional number of
 points in spheres, indeed show {\it scaling} properties on small
 scales $r <20$ Mpc/h, e.g., the former grows as a function of
 distance more slowly than the volume
 \citep{slmp98,hogg,2df_paper,sdss_paper,paper1}. This result implies
 that unconditional quantities are affected by systematic finite-size
 effects and thus do not give a reliable and meaningful estimation of
 correlations and amplitude of fluctuations.  In this situation the
 length scale $r_0$ can be an artifact of a statistical analysis,
 which assumes that the sample average is a meaningful estimation of
 the asymptotic density; i.e., it assumes that the distribution is
 homogeneous and that fluctuations have a small amplitude well inside
 the sample volume.

While estimations of real-space correlation properties can be affected
by finite-size effects, this is not the case when one counts galaxies
as a function of redshift or apparent magnitude. In this case indeed
one does not normalize statistical quantities to the sample average,
and large fluctuations have been found both in redshift
\citep{kerscher98,chiaki} and angular surveys
\citep[see][]{picard91,frith03}.  In particular, in a CCD survey of
bright galaxies within the northern and southern strips of the 2dF
Galaxy Redshift Survey (2dFGRS) conclusive evidence is found of
fluctuations of $ \sim 30\%$ in galaxy counts as a function
of apparent magnitude \citep{busswell03}.
Since in the angular region toward the southern galactic cap (SGC) a
deficiency, with respect to the northern galactic cap (NGC) in the
counts below magnitude $\sim 17$ (in the $B$ filter) was found,
persisting over the full area of the APM and APMBGC catalogs, this
would be evidence that there is a large void with a radius of about
$150$ Mpc/h, implying that there is more excess large-scale power than
detected in the 2dFGRS correlation function\footnote{These statistics,
  as mentioned above, normalize the amplitude of fluctuations to the
  estimation of the sample density.
From it the length scale $r_0 \approx 6-8$ Mpc/h is derived.}
\citep{norbergxi01,norbergxi02} or expected in the CDM models. It is
indeed evident that, because of the difference in the counts'
amplitude, and thus in the sample density, between the NGC and the SGC
samples, any estimation of the sample density is not stable.  Thus the
problems for the normalization of fluctuations amplitude to the
estimation of the sample density should be studied in great detail.

In this paper we use the 2dFGRS to study fluctuations in galaxy
distribution on large scales and to determine their statistical
properties in redshift and magnitude space. Our aim is to employ
statistical descriptors that do not, implicitly or explicitly, make
use of the normalization of fluctuations to the sample average, thus
avoiding the a-priori assumption of homogeneity inside a given sample.
Thus we determine conditional statistical properties, thereby
expanding our previous findings in this same survey
\citep{2df_paper,2df_2009}.  We find that the puzzle of the
coexistence of difference in densities in the NGC and SGC volumes on
large spatial scales with a relatively small typical length scale of a
few Mpc, can be understood as due to finite size effects in the
estimation of the correlation function.

The paper is organized as follows. In Sect.\ref{sec:samples} we
discuss the 2dFGRS data and the procedure used to construct the sub
samples for the statistical analysis. The methods for characterizing
homogeneous and heterogeneous distributions are briefly reviewed in
Sect.\ref{sec:statmeth}. In Sect.\ref{sec:results} we present the main
results of the analysis and consider in detail the problems related to
estimating fluctuations' amplitude normalized to the sample
density. The properties of fluctuations of the matter density field
predicted by theoretical models, in the linear regime, are briefly
reviewed in Sect.\ref{sec:mock}.  Then, to study expected nonlinear
fluctuations in standard theoretical models, we analyze the properties
of mock-galaxy catalogs, generated from the dark matter density fields
stemming from a cosmological N-body simulation --- the Millennium Run
by \citet{springel05}.  Finally in Sect.\ref{sec:discussion} we
discuss the results of the analysis of the 2dFGRS samples and outline
our main conclusions.


\section{The 2dFGRS samples}
\label{sec:samples}

The 2dFGRS \citep{colless01} measured redshifts for more than
$220,000$ galaxies in two strips in the SGC and in the NGC.  The
median redshift is $z\simeq0.1$. The apparent magnitude corrected for
galactic extinction in the $b_J$ filter is limited to 
$14.0<b_J<19.45$.  The selection of the samples used in the analysis
is described in detail in \citet{2df_paper}. Here we briefly
summarize the main points.

\begin{itemize} 
\item To avoid the effect of the irregular edges of the survey, we
  selected two rectangular regions: in the SGC there is a slice of
  size $84^\circ\times 9^\circ$ limited by
  $-33^\circ<\delta<-24^\circ$, $-32^\circ<\alpha<52^\circ$, while the
  NGC slice is smaller, i.e., $60^\circ\times 6^\circ$,  with
  limits $-4^\circ<\delta<2^\circ$, $150^\circ<\alpha<210^\circ$
  (coordinates are equatorial). The solid angles are $\Omega= 0.20 ,
  0.11$ steradians for the SGC and the NGC slices.

\item We selected galaxies in the redshift interval $0.01 \leq z \leq
  0.3$, with redshift quality parameter larger or equal to three, in
  order to get high quality redshifts \citep{hawkins03}.

\item We did not use a correction for the redshift-completeness mask
  and for the fiber collision effects. In fact, completeness varies
  mostly near the survey edges, which are excluded in our sample. We
  assumed that fiber collisions do not make a noticeable change in the
  small-scale correlation properties given that we set our lower
  cut-off to 0.5 Mpc/h,  which is larger than the 0.1 Mpc/h used by
  \citet{hawkins03}.

\item The metric distance is usually computed as in
  \citet{zehavietal02}:
\begin{equation}
\label{MetricDistance}
R(z) = \frac{c}{H_0}\int_{\frac{1}{1+z}}^{1}
{\frac{dy}{y \cdot \left(\Omega_M/y+\Omega_\Lambda
\cdot y^2 \right)^{1/2}}} \;,
\end{equation}
where we used the standard model parameters $\Omega_M=0.3$ and
$\Omega_\Lambda=0.7$, and $c$ is the light speed.   

\item The absolute magnitude was  computed as in
\citet{zehavietal02}:
\begin{equation}\label{AbsoluteMagnitude}
  M = b_J - 5 \cdot \log_{10}\left[R(z) \cdot (1+z)\right] - K(z) - 25\;,
\end{equation}
where $K(z)$ is the K-correction term \citep{hogg_kterm}

\item To calculate the K-correction $K(z)$ we used relations obtained
  by \citet{madgwick02} and we applied them as in \citet{2df_paper}.
 
\item The volume-limited (VL) samples were identified by two limits in
  the metric distance $R_{min} < R <R_{max}$ and two corresponding
  limits in the absolute magnitude $M_{min}$ and $M_{max}$.  In
  Table~\ref{tbl_VLSamplesProperties} we report the properties of the VL
  samples.
\end{itemize}
\begin{table}
\begin{center}
\begin{tabular}{|l|c|c|c|c|c|}
  \hline
  VL sample & $R_{min}$ & $R_{max}$ & $M_{min}$ & $M_{max}$ & $N_g$\\
  \hline
    SGC400 & 100 & 400 & -20.8 & -19.0 & 29373 \\
    NGC400 & 100 & 400 & -20.8 & -19.0 & 23208 \\
    SGC550 & 150 & 550 & -21.2 & -19.8 & 26289 \\
    NGC550 & 150 & 550 & -21.2 & -19.8 & 18030 \\
  \hline
\end{tabular}
\end{center}
\caption{Main properties of the obtained VL samples.  $R_{min}$,
  $R_{max}$ are the chosen limits for the metric distance; ${M_{min},
    \,M_{max}}$ are the corresponding limits in the absolute
  magnitude; $N_g$ is the number of galaxies in the sample. }
\label{tbl_VLSamplesProperties}
\end{table}

It is known that the completeness-mask for the survey includes
  the effects not only of redshift incompleteness (which varies both
  with field and with apparent magnitude), but also of the variations
  in the faint limiting magnitude across the survey regions.  In order
  to test the robustness of our results, without correcting directly
  for the incompleteness using ad-hoc assumptions, we  restricted
  the sample to the apparent magnitude limits $14.5<b_J<19.3$
  \citep{hawkins03}.  We report in
  Table~\ref{tbl_VLSamplesProperties2} the characteristics of the VL
  with these more conservative cuts.

 As a self-consistent test, we note below that the statistical
 analysis in the samples constructed with less conservative cuts agree
 with previous determinations for what concerns the galaxy counts as a
 function of apparent magnitude \citep{norberg01,busswell03}, the
 redshift distribution \citep{Ratcliffe98,busswell03}, and the standard
 two-point correlation function \citep{norbergxi01,norbergxi02}.

\begin{table}
\begin{center}
\begin{tabular}{|l|c|c|c|c|c|}
  \hline
  VL sample & $R_{min}$ & $R_{max}$ & $M_{min}$ & $M_{max}$ & $N_g$\\
  \hline
    SGC400 & 100 & 400 & -20.6 & -19.2 & 22872 \\
    NGC400 & 100 & 400 & -20.6 & -19.2 & 18407 \\
    SGC550 & 150 & 550 & -21.1 & -20.0 & 18890 \\
    NGC550 & 150 & 550 & -21.1 & -20.0 & 13495 \\
  \hline
\end{tabular}
\end{center}
\caption{As Table \ref{tbl_VLSamplesProperties} but for the case of
  more conservative cuts in apparent magnitudes; i.e. $14.5<b_J<19.3$,
  were used for selecting the galaxies}
\label{tbl_VLSamplesProperties2}
\end{table}


\section{Statistical methods}
\label{sec:statmeth}
In this section we review the main properties of stationary stochastic
point processes. These include both the ones that have a strictly
positive ensemble average density and those which have it equal to
zero.This discussion clarifies what the useful statistical methods are
in both cases for analyzing a {\it finite sample}. A more exhaustive
treatment can be found in \citet{book}.

\subsection{Volume average, ensemble average, and self-averaging property}

The problem of the statistical characterization of correlations and
fluctuations of a stochastic distribution of points in a finite sample
of volume $V$ can be rephrased as the problem of measuring
volume-averaged statistical quantities. The basic issue concerns
whether or not these are meaningful descriptors, i.e., whether or not
they give stable statistical estimations of ensemble averaged
quantities.  In this respect one has to consider various problems that
maybe clarified after the definition of the general probabilistic
properties of stochastic point processes.

First of all, we need to define the ensemble properties.  In general it
is assumed that galaxy distribution is a stationary stochastic
process, which means that it is statistically, translationally, and
rotationally invariant; i.e.,  it satisfies the condition of spatial
statistical isotropy and homogeneity in order to avoid special points
or directions\footnote{Because of the hypothesis of 
statistical stationarity, statistical quantities generally depend 
only of the scalar  distance between points.}. 
 Stationary stochastic distributions also satisfy these
conditions  when they have zero average density in the infinite
volume limit \citep{book}.

Due to the assumption of ergodicity (i.e., the ensemble average is
equal to the infinite volume average), the existence of a well-defined
average density implies that, for a single realization of the mass
distribution, the following limit is well-defined:
\begin{equation}      
 \label{shi1}     
 \lim_{R\rightarrow\infty}      
 \frac{1}{V(R;\vec{x}_0)}\int_{V(R,\vec{x}_0)}n(\vec{r})d\vec{r}    
=n_0     
\,\,\,, \forall \vec{x}_0\,. \,   
\end{equation}  
where $V(R,\vec{x}_0) \equiv 4\pi R^3/3$ is the volume of the sphere
of radius $R$, centered on an arbitrary point $\vec{x}_0$.  The
constant $n_0$ is strictly positive for homogeneous (or
uniform\footnote{To avoid confusion a stationary stochastic point
  process with a positive ensemble average density is sometimes
  denoted as uniform and nonuniform when this is equal to zero. Both
  are statistically homogeneous stochastic processes \citep{book}.})
distributions and zero for inhomogeneous ones (e.g. fractals).  It is
then clear that all statistical quantities that are normalized to
$n_0$ are meaningless when this is equal to zero in the ensemble
average or in the infinite volume limit.

Keeping these mathematical properties in mind we have to consider the
situation occurring when in a finite sample, of size $\sim V^{1/3}$,
the distribution is not homogeneous.  In this case the estimator of
the average mass density gives a large relative error with respect to
the ensemble value making it systematically biased.  This implies
that only statistical averages conditioned to the fact that the origin
of coordinates is a point of the set are well-defined.

Clearly a distribution can be inhomogeneous up to a length scale
$\lambda_0$\footnote{See below for a precise definition.} and
homogeneous for $r>\lambda_0$. Then for $r<\lambda_0$ the distribution
is characterized by large fluctuations, and the average density is not
a well-defined quantity if it is estimated in samples of size $V^{1/3}
< \lambda_0$. Instead density fluctuations are small for $r>\lambda_0$
and the sample density converges to the asymptotic (or ensemble) value
when the sample volume is such that $V^{1/3} > \lambda_0$.  The
precise behavior of this convergence is determined by the (weak)
two-point correlation properties of the distributions, and the
convergence will be slower when correlations are long-range
\citep{book}.


\subsection{Probability distribution of conditional fluctuations}

In \citet{paper1} we have introduced the scale-length (SL) analysis.
This consists in determining the number $N(r;R_i)$ of galaxies in
spheres of radius $r$, centered on the $i^{th}$ galaxy\footnote{This
  is thus a conditional quantity.}  whose distance from the origin is
$R_i$; that is,
\be
\label{eq1a} 
N(r;R_i) = \int_{V(r;R_i)} n(\vec{s}) d\vec{s} \;,
\ee 
where the integral is performed over the spherical volume $V(r;R_i)$
of radius $r$ centered on the $i^{th}$ galaxy at distance $R_i$ from
the origin, and $n(\vec{s})$ is the microscopic number density, which
is a function of the space position $\vec{s}$ .

When Eq.\ref{eq1a} is averaged over the whole sample, it gives an
estimate of the average conditional number of galaxies in spheres of
radius $r$
\be
\label{eq1} 
\overline{N(r)} = \frac{1}{M(r)} \sum_{i=1}^{M(r)} N(r;R_i) \;, 
\ee
where the sum is extended to the $M(r)$ galaxies, the $i^{th}$ at
radial distance $R_i$ from us, whose separation from the boundaries of
the sample is less than or equal to $r$.  In this way when $r$ grows
the number of values $M(r)$ over which the mean in Eq.\ref{eq1} is
performed decreases with $r$, because only those galaxies for which the
sphere is fully included in the sample volume are considered as
centers.

This estimator, known as the full-shell estimator
\citep{slmp98,kerscher,book}, has the advantage of making the weakest
a-priori assumptions about the properties of the distribution outside
the sample volume.  Indeed one may use incomplete spheres by counting
the number of galaxies inside a portion of a sphere and weighting this
for the corresponding volume of the spherical portion
\citep{kerscher}.  However, this method implicitly uses the assumption
that what is inside the incomplete sphere is a statistically
meaningful estimate of the distribution in the whole spherical
volume. This is incorrect when the distribution presents large
fluctuations. For example in the part of a spherical volume that lies
outside the sample boundaries, there can be a void or a large-scale
structure and in this situation the weighted estimation is biased
\citep{book}.  When the full-shell estimator is used, one should
consider that there is an intrinsic selection effect related to the
geometry of the samples, which are small portions of spheres. When $r$
is large only the more distant part of the sample is explored by the
volume average \citep{book,paper1}. Indeed, for large-sphere radii,
$M(r)$ decreases and the location of the galaxies contributing to the
average in Eq.\ref{eq1} is mostly placed at radial distance in the
range $\sim [R_{min}+r$, $R_{max}-r]$ from the radial boundaries of
the sample at $[R_{min}$, $R_{max}]$. Given the geometry
of the samples for large $r$,  galaxies contributing to $M(r)$ will
also lie toward the center of the spherical portion.

When Eq.\ref{eq1} scales as 
\be
\label{eq2} 
 \overline{N(r)} \sim r^D 
\ee 
and $D=3$, the distribution is homogeneous, while for \mbox{$D<3$} it has
long-range power-law correlations \citep{book}.  The scaling of
Eq.\ref{eq2} with $D<3$ can be interpreted as the signature that the
distribution is a fractal; however there are point distributions
that, by construction, are not fractal objects but which may exhibit
a scaling of the type given by Eq.\ref{eq2}
\citep[see][]{book}.

From Eq.\ref{eq2} we obtain that, in general, for $D<3$
the conditional density scales as
\[
\overline{n(r)} = \frac{\overline{N(r)}}{V(r)}\sim r^{D-3} \,;
\]
 in this situation the average density is not a well defined quantity
 and the sample density is depends on the sample size; i.e., it does
 not give a meaningful estimation of the ensemble average density.

When a distribution is fractal (or generally inhomogeneous) on small
scales and homogeneous on large scales, then we can identify the
homogeneity scale $\lambda_0$ to be the scale such that \citep{book}
\be
\label{eq2a} 
 \overline{N(r)} \sim r^3 \;\;,\; \forall r > \lambda_0 \;.  
 \ee
Depending on the details of the crossover from the strongly correlated
regime at $r<\lambda_0$ to the weakly correlated one at $r>\lambda_0$
different definitions of $\lambda_0$ can be adopted, but they all
satisfy the condition given by Eq.\ref{eq2a}. For instance one may
define such a scale to be the one at which the rms fluctuation have
twice the value of the average density.  If strong clustering occurs
only for $r<\lambda_0<V^{1/3}$ it is required to make the statistical
analysis suitable to describe large fluctuations in this range of
scales. Instead for $\lambda_0<r<V^{1/3}$ usual unconditional
quantities are well-defined and their convergence to the ensemble
values can be studied.


The estimator defined by Eq.\ref{eq1} gives the first moment (i.e., the
average number of points in spheres of radius $r$) of the PDF of
conditional fluctuations $f(N;r)$ computed, at fixed $r$, from the
values $\{N(r;R_i)\}_{i=1,..,M(r)}$.  This is generally different from
the PDF of unconditional fluctuations --- considered by, e.g.,
\citet{saslaw} --- both for homogeneous and inhomogeneous
distributions, the difference being more important in the former case.

When a distribution becomes homogeneous; i.e.,  Eq.\ref{eq2a} is
satisfied, the PDF is expected to converge in a finite volume to a
Gaussian function\footnote{This is clearly the case if the number of
  points is large enough, otherwise the PDF is described by the
  Poisson distribution} \citep{book}; i.e.,
 \be
\label{pdf}
f(N;r \gg \lambda_0) \simeq \frac{1} {\sqrt{2 \pi
    \overline{\Sigma^2(r)}}} \exp \left( - \frac{[N(r) -
  \overline{N(r)}]^2} {2 \overline{ \Sigma^2(r)}} \right) \;,
\ee 
where $\overline{\Sigma^2(r)}$ is the estimation of the variance of
the random variable $N(r)$ (see below).

The second moment of $f(N,V)$ gives the conditional variance. For
inhomogeneous distributions, this is such that
\be
\label{second_moment}
\overline{\delta(r)^2} \equiv \frac{\overline{\Sigma^2}(r) } {
  \overline{N(r)}^2} = \frac{ \overline{N(r)^2} - \overline{N(r)}^2}{
  \overline{N(r)}^2} \sim 1 \;, 
\ee 
where the last equality means that fluctuations are persistent
\citep{gsl00}.  On the other hand, for homogeneous distributions with
any kind of small-amplitude correlations, we find that \citep{gsl00}
\be 
\overline{\delta(r)^2} \ll 1 \;.  
\ee 
When the sample density $n_S$ is well-defined; i.e.,  it does not
depend on the sample volume, one may define the reduced two-point
correlation function that can be written as \citep{book}
\be 
\label{xi1}
\xi(r) = \frac{1}{4\pi r^2} \frac{d\overline{N(r)}} {dr} \frac{1}{n_S}
-1 \;.  \ee
In this case the homogeneity scale can be, for instance, defined as
the scale beyond which $\xi(r) <1$, or the scale such that
$\delta^2(\lambda_0) =1$ \citep[see][]{book}.


\section{Results from the 2dFGRS} 
\label{sec:results}

In this section we present the results of the statistical analysis of
the VL samples of the 2dFGRS catalog discussed in
Sect.\ref{sec:samples}.  We start by presenting the SL analysis to
then move to the description of the determination of the PDF of
conditional fluctuations. Furthermore, we consider its first moment;
i.e., the average number of points in spheres. To illustrate the
usefulness of the SL analysis, we consider its relation to the counts
of galaxies as a function of radial distance and of apparent
magnitude.  This allows us to discuss in detail the relation between
small scale two-point correlations and large scale properties of
fluctuations in the galaxy density field.  We then consider the
finite-size effects that systematically affect the determination of
fluctuations amplitude normalized to the sample density. In addition
we compute the average of the SL determinations
$\{N(r;R_i)\}_{i=1..M(r)}$ in bins of radial distance.  The comparison
of the behaviors in the NGC and SGC slices allows us to place a lower
limit on the homogeneity scale.


\subsection{The scale-length analysis}

In Figs.\ref{fig_ngp400_sl}-\ref{fig_sgp550_sl} the behavior of the SL
analysis is shown in the four 2dFGRS samples we considered. One may
note that in all cases there are large density fluctuations in the
correspondence of the location of galaxy large-scale structures. In
Fig.\ref{fig_sgp550_color_sl} a three-dimensional plot is shown of the
same SL analysis reported in Fig.\ref{fig_sgp550_sl}.  One may see how
well structures are identified by this analysis.

The number of points $M(r)$ over which $N(r;R_i)$ is computed, as a
function of the sphere radius $r$ is shown in Fig.\ref{fig_cs}: when
the sphere radius $r$ gets larger, the $M(r)$ decreases quite
rapidly. This is due to the geometrical selection effect previously
discussed.  In addition, the solid angle of the SGC slice is twice
that of the NGC slice, and thus for $r>20$ Mpc/h, there are more
center points in the SGC samples than in the NGC ones.

Let us briefly discuss the main features that we detect in the various
samples \footnote{ Note that, for instance, \citet{eke04,einasto} used
  different methods to identify the same structures we observe.}
\begin{figure}
\begin{center}
\includegraphics*[angle=0, width=0.5\textwidth]{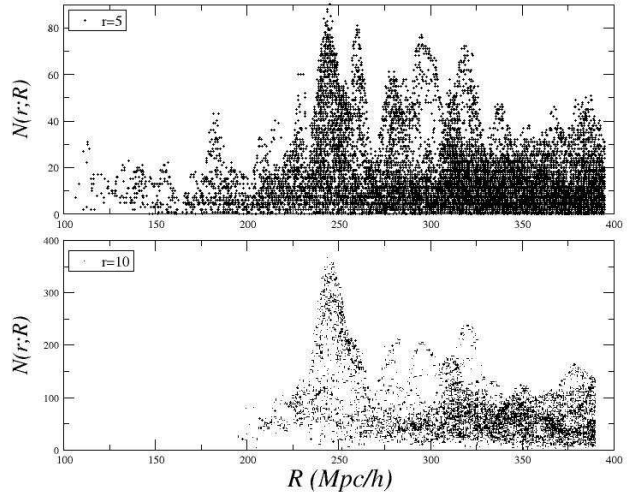}
\end{center}
\caption{ From top to bottom the SL analysis for the sample NGC400
  with $r=5,10$ Mpc/h.  }
\label{fig_ngp400_sl}
\end{figure}
\begin{figure}
\begin{center}
\includegraphics*[angle=0, width=0.5\textwidth]{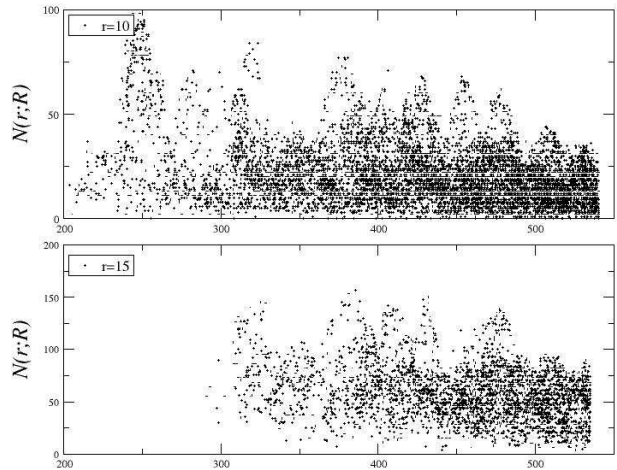} 
\end{center}
\caption{The same as Fig.\ref{fig_ngp400_sl} but now for the 
  sample NGC550 with $r=10,15$ Mpc/h.}
\label{fig_ngp550_sl}
\end{figure}
\begin{figure}
\begin{center}
\includegraphics*[angle=0, width=0.5\textwidth]{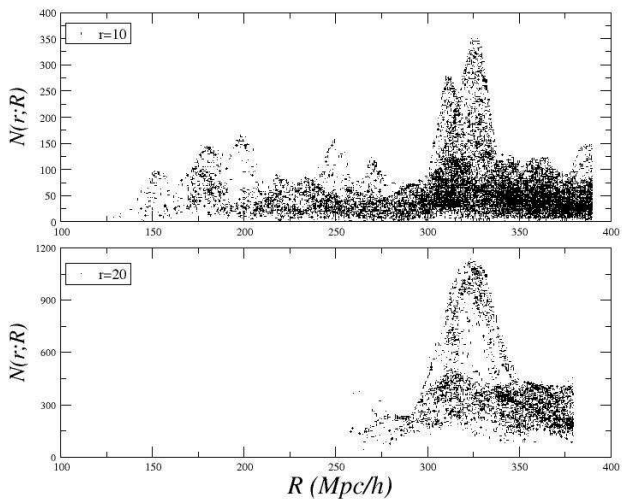}
\end{center}
\caption{The same as Fig.\ref{fig_ngp400_sl} but now for the 
  sample SGC400 with $r=5,10,20$ Mpc/h.  }
\label{fig_sgp400_sl}
\end{figure}
\begin{figure}
\begin{center}
\includegraphics*[angle=0, width=0.5\textwidth]{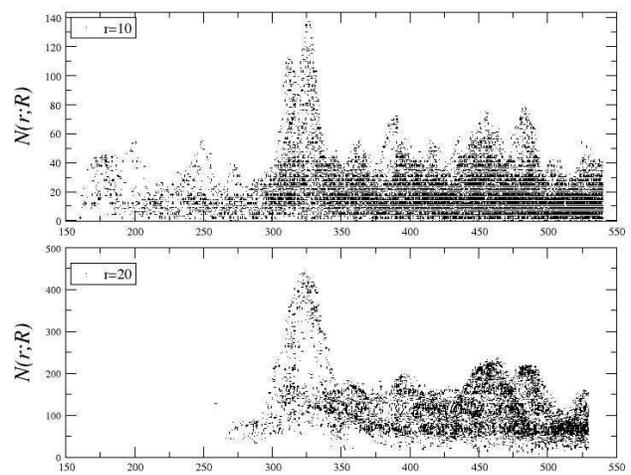}
\end{center}
\caption{The same as Fig.\ref{fig_ngp400_sl} but now for the 
  sample SGC550 with $r=10,20,30$ Mpc/h.}
\label{fig_sgp550_sl}
\end{figure}
\begin{figure*}
\begin{center}
\includegraphics*[angle=0, width=17cm]{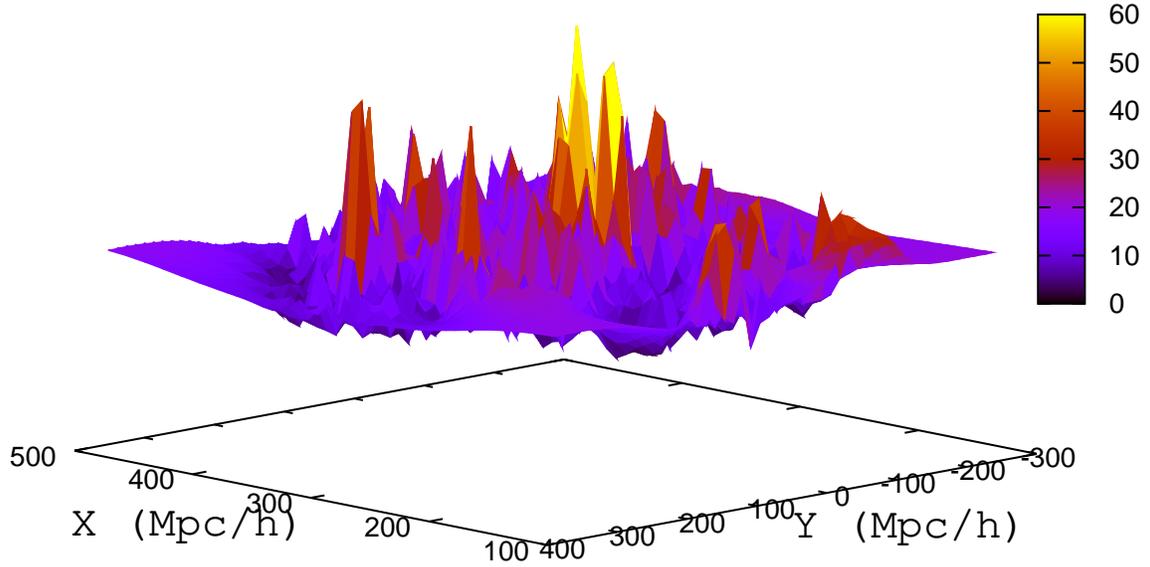}
\end{center}
\caption{The SL analysis for the SGC550 sample.  On the $X$ and $Y$
  axes the coordinate of the center of a sphere of radius $r=10$ Mpc/h
  (centered on a galaxy) is reported and on the $Z$ axis the number of
  galaxies inside it.  The mean thickness of this slice is about 50
  Mpc/h.  Large fluctuations in the density field traced by the SL
  analysis are located in the correspondence of large-scale
  structures.  }
\label{fig_sgp550_color_sl}
\end{figure*}

\begin{figure}
\begin{center}
\includegraphics*[angle=0, width=0.5\textwidth]{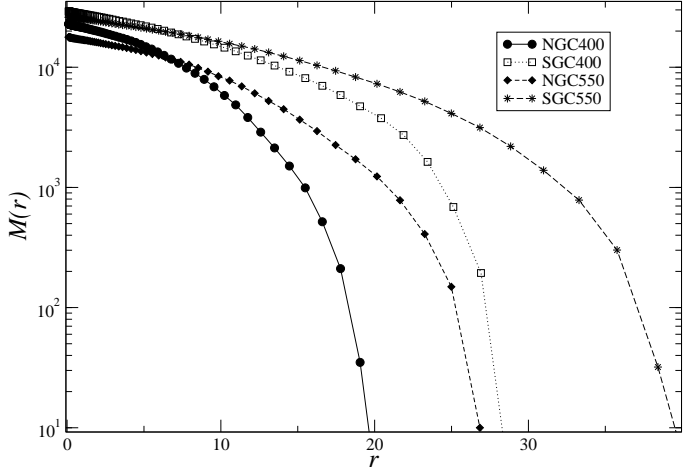}
\end{center}
\caption{Number of center-points $M(r)$ as a function of the sphere
  radius $r$ in the various samples considered.}
\label{fig_cs}
\end{figure}

\begin{itemize}
\item NGC400: there are large structures, which transversely cross the
  sample, at about $240$ Mpc/h and at 260, 270, 290 and 320 Mpc/h.
  All have approximately the same thickness of about 30-40 Mpc/h.
  When the sphere radius is increased to $r=$ 10 Mpc/h the most
  prominent structure remains the one at about $250$ Mpc/h, which is
  not sampled anymore when $r=15$ Mpc/h. This is due to the sphere
  centers being located toward the faraway boundaries of the sample,
  because of the geometrical selection effect discussed previously.
  For $r=15$ Mpc/h, although only a few points (i.e $M \approx$ 3000)
  are effectively considered as sphere centers, density fluctuations
  are still large, determining variation of a factor four in the
  determination of the number of points in spheres at different radial
  distances.

\item NGC550: the structure at $R\approx 250$ Mpc/h is clearly visible
even in this sample, and other structures are present at larger 
radial distances $R>350$. Mpc/h. Fluctuations are still large up
to the largest sphere radius $r=25$ Mpc/h where however 
the number of centers rapidly decreases for the geometrical 
selection effect discussed in Sect.3

\item SGC400: for sphere radius $r=10$ Mpc/h the situation is
  similar to the NGC400 case, except for the fact that the radial
  distances corresponding to the large variations in the density field
  are different.  This shows that large-scale structures and the
  corresponding large fluctuations detected by the SL analysis are
  quite typical of the galaxy distribution. There are two prominent
  large-scale structures at radial distance of the order of $R\sim$
  320 Mpc/h. Finally for $r=20$ Mpc/h the sample is
  dominated by one of the two structures just mentioned, which corresponds
  to a variation of order five in $N(r;R)$.

\item SGC550: a structure at $R\approx 320$ Mpc/h with thickness of
  about 40 Mpc/h is present inside this sample as well.  In addition
  other structures are visible for $R>400$ Mpc/h. For the largest
  sphere radius $r=30$ Mpc/h, where only a few points are considered as
  sphere centers, fluctuations in $N(r;R)$ are of order five and they
  are due to structures located at radial distances $R>380$ Mpc/h.
\end{itemize}

In summary galaxy distribution in these samples is characterized by
several large scale structures which cross their volumes.  These
structures are typical and persistent; i.e., they are detected at
different radial distances and in the two different directions toward
the SGC and NGC. They correspond in the SL analysis to large
fluctuations in the density field: 
the thickness of the structures is partially an intrinsic property and
partially due to the 'Finger of God' effect --- the redshift space
extension of large clusters and super-clusters --- convolved with the
sphere size used to measure the conditional density.  Typical velocity
dispersions are up to $\sim 10^3$km/s, although \citet{eke04} conclude
that the typical velocity dispersion for 2dF groups and clusters is
260 km/s, which only corresponds to 2.6 Mpc/h in an additional radial
extension. Thus while peculiar velocities can increase the size of
radial structures, their effect cannot explain the whole range of
distances which characterize the observed structures.

The maximum allowed sphere radius we considered, which, as discussed,
is set by the geometry of the samples, is $r=30$ Mpc/h.  For this
value of $r$ we find fluctuations of order four in $N(r;R)$ and this
allows us to conclude that the homogeneity scale is certainly larger
than $40$ Mpc/h. In what follows we reinforce this conclusion by
considering suitable statistical measurements.


\subsection{Probability distributions of conditional fluctuations}

We now turn to the discussion of the PDF of conditional fluctuations
The behaviors in the various samples are shown in
Figs.\ref{fig_ngp400_pdf}-\ref{fig_sgp550_pdf}. We limit the
discussion to the case where the number of determinations $\{ N(r,R_i)
\}_{i=1,..,M(r)}$ at fixed sphere radius $r$ is larger than few
thousands points. For smaller numbers the measurement is affected by
weak statistics and by finite-size effects, thus not leading to a
statistically robust result.  As discussed previously, when the sphere
radius increases, there is a decrease in the number of centers, and
thus a for large $r$, whose precise value is determined by the
geometry of each  specific sample, the measurement is affected by
large statistical and systematic effects (see Fig.\ref{fig_cs}).

The first point to note is that in all cases the maximum of $f(N,r)$
is statistically stable; i.e.,  it does not change when it is computed
in the whole sample or in two non-overlapping sub samples with equal
volume (each half of the sample volume) at small (sub-sample $S_1$)
and at large (sub-sample $S_2$) radial distance (see
Fig.\ref{fig_ngp400_pdf} and Fig.\ref{fig_sgp400_pdf}).

The tail for large values of $N$ is instead affected by the different
fluctuations which are present in different sub-volumes. The trend is
obvious: the larger the fluctuations of $N(r;R)$ the more extended
toward high $N$ values is the tail of $f(N,r)$. In the deepest
samples, e.g. SGC550, there is a single structure that dominates the
distribution. However this is placed in the middle of the sample and
for this reason, apparently, there is no a systematic difference in
the PDF when this is computed in the two half volumes of the sample,
one nearby and one far-away.  However, it is clear that in this
situation the shape of the PDF will be strongly affected by this
fluctuation. Indeed, in each sub sample, for the largest sphere radius
$r$ we find that $f(N,r)$ is systematically distorted with respect to
smaller sphere radii. This is because the volume average cannot explore
 the full sample properly because of the geometrical selection effect
which, as discussed, is present in the determination of $N(r;R)$. To
properly determine the PDF on scales $r>20$ Mpc/h larger samples are
thus required.

In all cases the PDF is systematically different from a Gaussian
function, except for the case of NGC550 for which there are the
weakest statistics, and it is characterized by a long $N$ tail which
is directly related to the large scale structures present in these
samples. In addition the PDF differs in different samples, especially
for $r>10$ Mpc/h. This implies that, because of the weak statistics
and small volumes, a clean determination of the PDF is impossible. For
this reason we limit our discussion in what follows to the first
moment of the PDF, leaving the determination of the second moment to
the other samples. for instance those of the SDSS, where spatial
volumes will be larger \citep{paper1}.

\begin{figure}
\begin{center}
\includegraphics*[angle=0, width=0.5\textwidth]{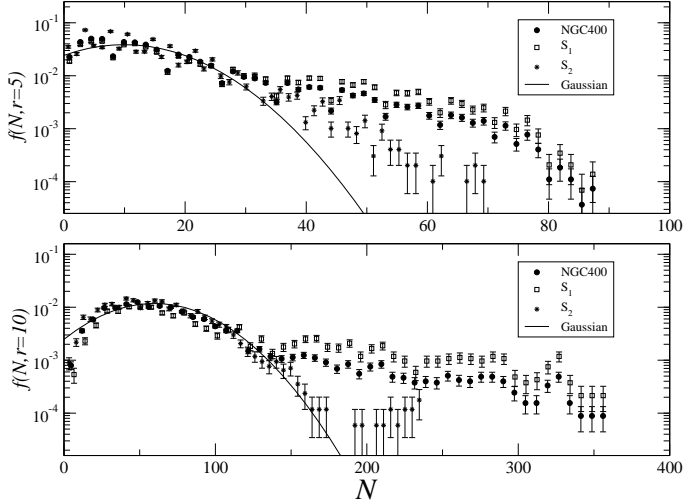}
\end{center}
\caption{Probability density function $f(N,r)$ of the values
  $\{N(r;R_i)\}_{i=1,..,M(r)}$ for NGC400 in the whole sample and the
  best fit with a Gaussian function.  The different sphere radius $r$
  is reported on the y-axis. The behavior in two non-overlapping
  sub samples (labeled as $S_1$ and $S_2$ is also reported --- see
  text for details. Poisson error bars are shown as a reference.}
  \label{fig_ngp400_pdf}
\end{figure}
\begin{figure}
\begin{center}
\includegraphics*[angle=0, width=0.5\textwidth]{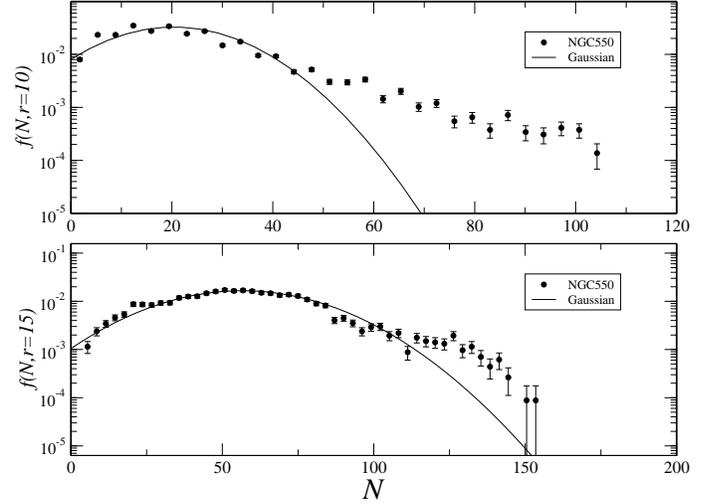}
\end{center}
\caption{The same as in Fig.\ref{fig_ngp400_pdf}  for the sample
  NGC550 with $r=10,15$ Mpc/h. For $r=15$ Mpc/h there are only 3765
  determinations.}
\label{fig_ngp550_pdf}
\end{figure}
\begin{figure}
\begin{center}
\includegraphics*[angle=0, width=0.5\textwidth]{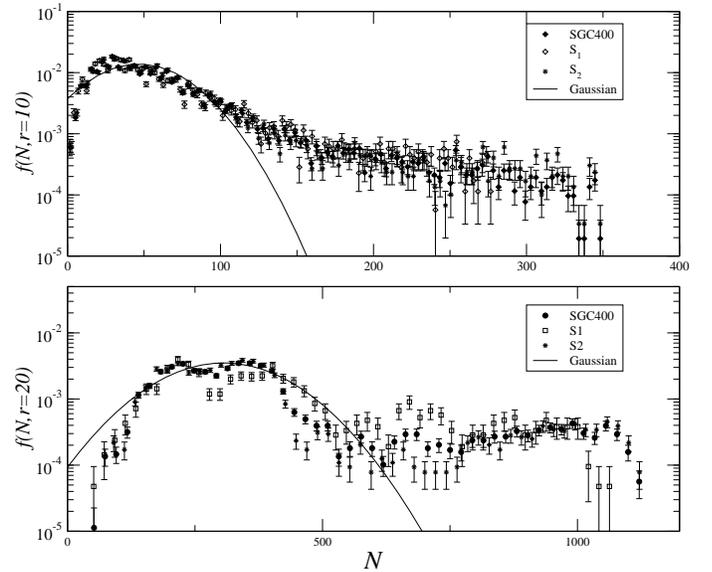}
\end{center}
\caption{The same as in Fig.\ref{fig_ngp400_pdf} for the 
  sample SGC400 with $r=5,10,20$ Mpc/h.  }
\label{fig_sgp400_pdf}
\end{figure}
\begin{figure}
\begin{center}
\includegraphics*[angle=0, width=0.5\textwidth]{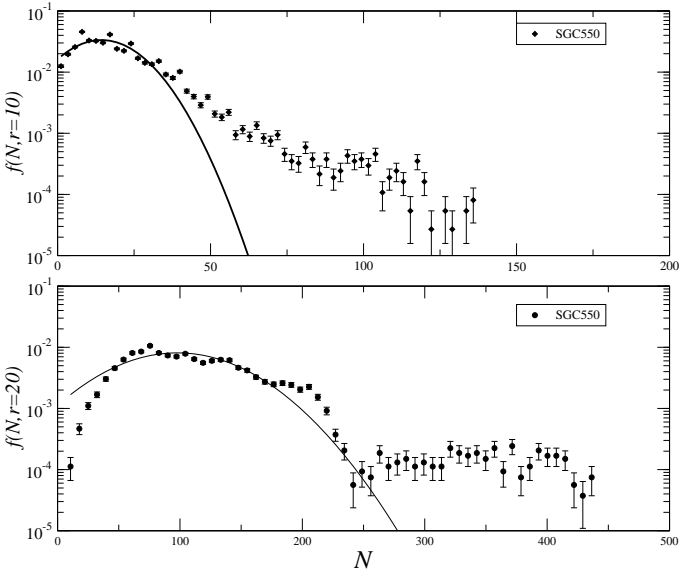}
\end{center}
\caption{The same as in Fig.\ref{fig_ngp400_pdf} for the sample SGC550
  with $r=10,20$ Mpc/h. }
\label{fig_sgp550_pdf}
\end{figure}

In Fig.\ref{fig_pdf_fit} we show the collapse plot of the PDF in the
various samples and for the different sphere radius considered. A
rough fit of the large $N$ tail is given by
\[
f(N,r) \sim N^{-3} \;.
\]
However, this result requires better samples before being confirmed.

\begin{figure}
\begin{center}
\includegraphics*[angle=0, width=0.5\textwidth]{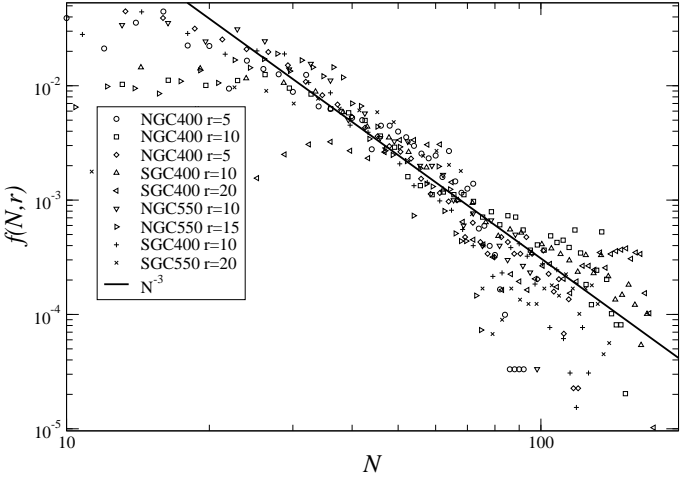}
\end{center}
\caption{Collapse plot of the $f(N,r)$ in the various samples and for
  the different sphere radius considered. The normalization of the
  different samples has been performed in an arbitrary way and on the
  X-axis there are arbitrary units.}
\label{fig_pdf_fit}
\end{figure}


\subsection{Conditional average number of galaxies in spheres}

In Fig.\ref{fig_NR} we show the determination of the whole-sample, average
conditional number of points in spheres; i.e., Eq.\ref{eq1}. Given that
the PDF in all samples are statistically stable; i.e., they do not show
important systematic differences in different sub-volumes, the full
sample volume average provides a meaningful statistical
quantity. These behaviors are the same as the ones found by
\citet{2df_paper} for the average conditional density.  We find that
in all samples,
\be
\label{nr_est} 
\overline{N(r)} = B r^D \;,
\ee 
with $D=2.2\pm 0.1$ and $B$ a constant.  This result is statistically
robust up to $r=30$ Mpc/h although for the SGC550 sample the analysis
can reach, with a weaker statistics, $r\sim$ 40 Mpc/h.  In all cases,
however, the determinations for scales $r>20$ Mpc/h are done on a
relatively small number of points (see Fig.\ref{fig_cs}), so they can
be subject to systematic fluctuations. The amplitude between the
NGC400 and SGC400 samples (which have higher number density than that
of the deeper samples) differs for about $\sim 30\%$ (see
Fig.\ref{fig_NR2}). This is due to the different fluctuations, which
are present in the different volumes, and this implies that the
samples are not large enough to precisely determine the
amplitude. This difference is reflected in the number counts as a
function of apparent magnitude as we discuss below.  In the NGC550 and
SGC550 samples, although the samples volume is larger, the constant
$B$ in Eq.\ref{nr_est} varies $\sim 20 \%$ (see Fig.\ref{fig_NR2}).

To briefly discuss the determination of the constant pre-factor $B$ in
Eq.\ref{nr_est} \citep{jsl,book} and the normalization of the
conditional average density in different VL samples one should
consider the joint conditional probability of finding a galaxy of
luminosity $L$ at distance $\vec{r}$ from another galaxy; i.e.,  the
(ensemble) conditional average number of galaxies $\langle
\nu(L,\vec{r}) \rangle_p d\vec{r}dL$ with luminosity in the range
$[L,L+dL]$ and in the volume element $d\vec{r}$ at distance $r$ from
an observer located on a galaxy.  We can make the greatly simplifying
assumption that
\be
\label{gal2}   
\langle \nu(\vec{y},L) \rangle_p = 
\phi(L) \times \langle n(\vec{y}) \rangle_p \;,
\ee   
where $\langle n(\vec{y}) \rangle_p$ is the average conditional
density and $\phi(L)$  the luminosity function, such that $\phi(L)dL$
gives the probability that a randomly chosen galaxy has luminosity in
the range $[L,L+dL]$.  
Although there is clear evidence that there is a correlation between
them and, for instance, that the brightest elliptical galaxies are
found in the center of rich galaxy clusters, it has been tested that
this is nevertheless a reasonable assumption in the galaxy catalogs
available so far.  To properly treat the case where the correlation
between galaxy positions and luminosities is taken into account, one
would need to use the multi-fractal formalism. This goes beyond the
scope of the present manuscript \citep[see][]{book}. An additional,
much stronger assumption often adopted in Eq.\ref{gal2} is that the
space density is a constant; i.e.,  $\langle n( \vec{r} ) \rangle =
const.$ This assumption is for instance the basis of the so-called
standard minimum variance estimator \citep{davishuchra}.  It is clear
that we avoid making this further assumption because we want to test
whether the space density is (or can be approximated by) a simple
constant.

Using Eq.\ref{gal2} we may write the
conditional average number of galaxies as a function of distance (in
case $D=$const.)
\bea
\label{gal11}  
&&\langle N (r; L^1_{VL}<L<L^2_{VL}) \rangle_p  
\\
\nonumber   
&&
= \int_0^r \int_{L^1_{VL}}^{L^2_{VL}} 
\langle \nu(\vec{y},L) \rangle_p dL d\vec{y}
\\
\nonumber   
&&= B r^D  \int_{L^1_{VL}}^{L^2_{VL}} 
\langle \phi(L) \rangle_p dL= B_{VL}r^D  \;,
\eea 
where $N(r;L^1_{VL}<L<L^2_{VL})$ is the number of galaxies in a
sphere of radius $r$ and with intrinsic luminosity in the range
$[L^1_{VL},L^2_{VL}]$, and $B_{VL}$ is the amplitude of the number
counts in the VL sample with these limits in absolute luminosity.
Because the luminosity function has an exponential cut-off at $L^*$,
VL samples containing brighter galaxies show a smaller $B_{VL}$. By
knowing the shape of the luminosity function, it is simple to normalize
the different $B_{VL}$ in different VL samples \citep{jsl}.

 \begin{figure}
\begin{center}
\includegraphics*[angle=0, width=0.5\textwidth]{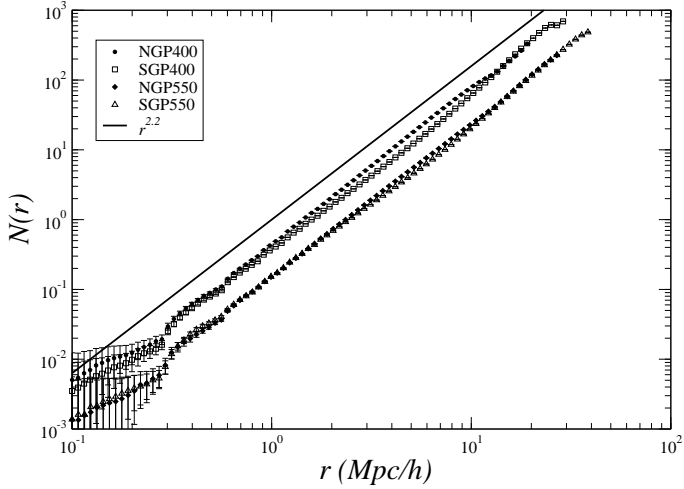}
\end{center}
\caption{Average number of points in spheres of radius $r$ around a
  galaxy. The difference amplitude in samples with different limits in
  absolute magnitude is simply ascribed by the effect of the
  luminosity function (see text). Error bars are estimated 
  by the sample dispersion on the average value.}
\label{fig_NR}
\end{figure}
 \begin{figure}
\begin{center}
\includegraphics*[angle=0, width=0.5\textwidth]{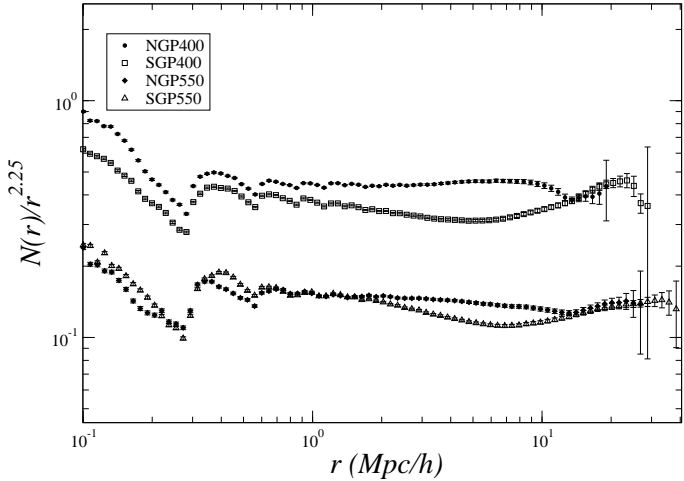}
\end{center}
\caption{The same as in Fig.\ref{fig_NR} but divided by the best-fit
  power-law behavior $r^{2.25}$. The variation in the amplitude $B$ in
  Eq.\ref{nr_est} is  clearer in this representation.  The
  determination for $r>20$ Mpc/h is  subject to systematic
  fluctuations, due to the limited volume and the weaker statistics.
}
\label{fig_NR2}
\end{figure}
%


\subsection{Average in radial bins}

We can now use the data obtained by the SL method to investigate
whether there is a convergence to homogeneity on some large scales
$r>30 \div 40$ Mpc/h; i.e., the largest sphere radius allowed by the
geometry of the samples.  This is done as follows. We divide the whole
range of radial distances in bins of thickness $\Delta R$ and 
compute the average
\be
\label{NRave}
\overline{N(r;R,\Delta R)}
 = \frac{1}{M_b} 
\sum_{R_j\in [R,\Delta R] }^{j=1,M_b} N(r; R_j)\;,
\ee
where the sum is extended to the $M_b$ determinations of $N(r;R)$, such
that the radial distance is in the interval range $[R,R+ \Delta
  R]$. It is clear that for this measurement we have to use a small
$r$ to avoid overlapping in space between neighboring bins in
radial distance.  The variance on the mean is
\be
\label{NRaveerr}
\overline{\Sigma(r; R,\Delta R)^2} = 
\sum_{R_j\in [R,\Delta R] }^{j=1,M_b}
\frac{\left(N(r; R_j) - 
\overline{N(r;R,\Delta R)} \right)^2}{M_b(M_b-1)} \;.
\ee
The error analysis in Eq.\ref{NRaveerr} assumes that the $N(r; R_j)$
are independent, but in fact they are correlated. The errors on these
points are therefore substantially under-estimated. However if there
is a trend toward homogenization, the error caused by neglecting this
correlation will be smaller hence the under estimate of the error
bars. Only for the case of a highly correlated distribution does 
Eq.\ref{NRaveerr} underestimate the error bars, which than represent
a lower limit of the ``true'' error bars.

The quantity  given by Eq.\ref{NRave} and its error (Eq.\ref{NRaveerr})
provide an estimation of the number of points in spheres of radius $r$ 
averaged in  thickness bin $\Delta R$. We expect that, if the 
distribution converges to uniformity on a scale $\lambda_0$,   
then correspondingly $\overline{N(r;R,\Delta R>\lambda_0)}$ does not
show large fluctuations as a function of $R$.

Results for the four samples are shown in
Figs.\ref{fig_NR_ave_400}-\ref{fig_NR_ave_550}.  One may note that, for
the largest radial bin chosen $\Delta R=75$ Mpc/h, there is no trend
in homogenization, but instead  the measurements in bins centered on 
different $R$ wildly scatters; i.e., their values are outside the
statistical error bars given by Eq.\ref{NRaveerr}. This shows that
large-scale structures have an amplitude that is incompatible with
homogeneity on scales smaller than $\lambda_0 = \Delta R = 75$ Mpc/h.
\begin{figure}
\begin{center}
\includegraphics*[angle=0, width=0.5\textwidth]{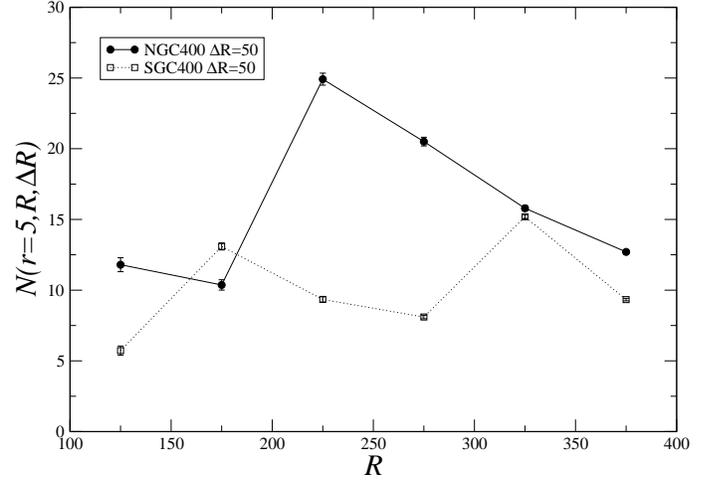}
\end{center}
\caption{Behavior of  Eq.\ref{NRave} with $r=5$ Mpc/h for NGC400
  and SGC400 and $\Delta R=50$ Mpc/h.  The error is computed through
  Eq.\ref{NRaveerr}.  }
\label{fig_NR_ave_400}
\end{figure}
 \begin{figure}
\begin{center}
\includegraphics*[angle=0, width=0.5\textwidth]{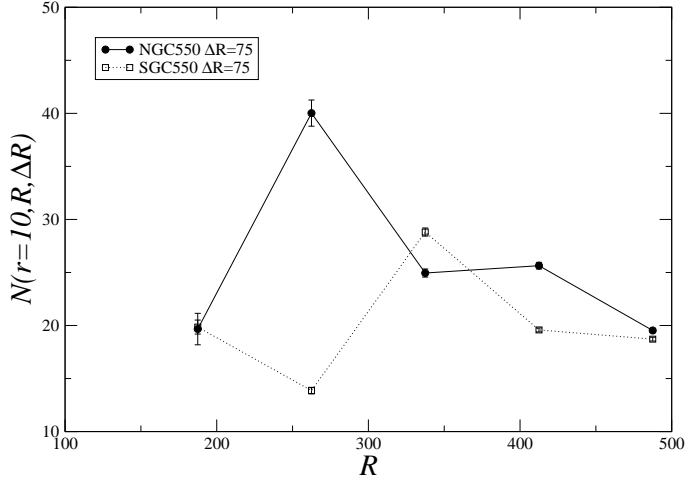}
\end{center}
\caption{As in Fig.\ref{fig_NR_ave_400} but for NGC550 and SGC550 with
  $r=10$ Mpc/h and $\Delta R=50$ Mpc/h.}
\label{fig_NR_ave_550}
\end{figure}

It is interesting to note that the large fluctuations between the NGC
and SGC samples cannot be generated by some redshift-dependent
effect, such as the inclusion of galaxy evolution in the computation
of absolute magnitudes. Indeed, such a correction, which is expected to
be  small anyway, given that the redshifts involved do not exceed 0.2,
would affect  both samples in the same way. Thus by comparing the
estimation of the density in bins in the same range of radial
distances, we can conclude that the fluctuations we have detected are
intrinsic to the distribution of galaxies in these samples. A similar
argument can be made for the effect of different cosmologies in the
computation of the metric distance.


\subsection{Radial counts in VL samples}

A complementary way to study fluctuations on large scales in galaxy
redshift surveys is represented by the determination of the radial
counts; i.e., the counts of galaxies as a function of the radial
distance in VL samples \citep{gsl00}. In order to have a statistical
estimator and to evaluate fluctuations, we divided the angular area of
the samples into $N_f=20$ non overlapping sub fields of equal solid
angle.  For each we compute the differential radial density
$n_i(R;\Delta R)$ in bins of thickness $\Delta R =10$ Mpc/h , where
$i^{th}$ labels the sub field and $R$ is, as usual, the radial
distance. We then can compute the average
\be
\label{diff1}
\overline{n(R;\Delta R)} = \frac{1}{N_f} \sum_{i=1}^{N_f} n_i(R;\Delta R)
\ee
and the sample variance
\be
\label{diff2}
\overline{\sigma^2_n(R;\Delta R)} = \frac{1}{N_f-1} \sum_{i=1}^{N_f} 
(n_i(R;\Delta R) - \overline{n(R;\Delta R)})^2 \;.
\ee
Results are shown in Figs.\ref{fig_counts_ngp400}-\ref{fig_counts_sgp550}.
\begin{figure}
\begin{center}
\includegraphics*[angle=0, width=0.5\textwidth]{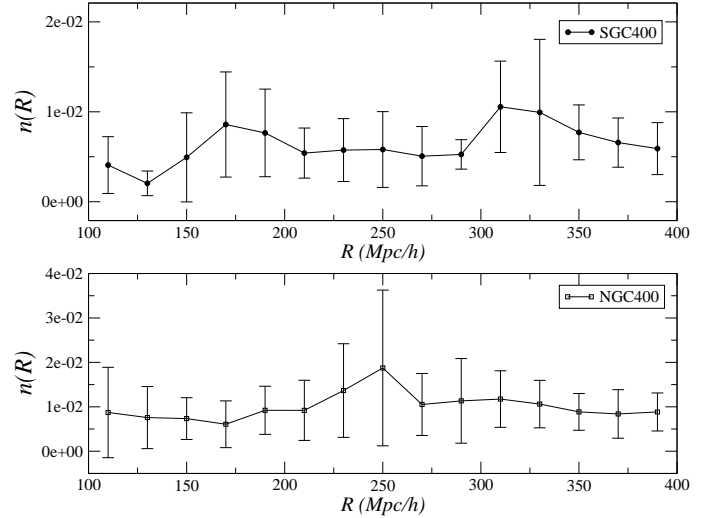}
\end{center}
\caption{The average differential radial density 
$\overline{n(R;\Delta R)}$ (Eq.\ref{diff1}) and its error (Eq.\ref{diff2}) 
 for the samples NGC400 and SGC400.}
\label{fig_counts_ngp400}
\end{figure}
\begin{figure}
\begin{center}
\includegraphics*[angle=0, width=0.5\textwidth]{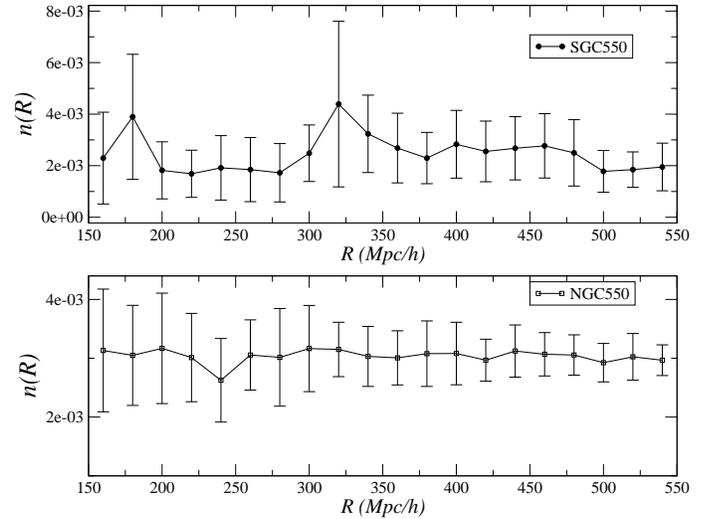}
\end{center}
\caption{The same as in  Fig.\ref{fig_counts_ngp400} for SGC550 and NGC550.}
\label{fig_counts_sgp550}
\end{figure}
In the NGC400 sample the structure at 250 Mpc/h is visible as a
relatively large fluctuation of $\overline{n(R;\Delta R)}$ and a
correspondingly large error. This means that this structure partially
covers the angular area of the survey. In the NGC550 sample, the
radial density is flatter, although there is a large dispersion.  In
the SGC400 sample, the structures at 160 Mpc/h and 320 Mpc/h are
identified as local enhancements of $\overline{n(R;\Delta R)}$. The
same occurs for the SGC550 case, where the same two structures are
visible. By comparing
Figs.\ref{fig_counts_ngp400}-\ref{fig_counts_sgp550} with
Figs.\ref{fig_ngp400_sl}-\ref{fig_sgp550_sl} one may note that the SL
analysis is a much more powerful method than the simple counting as a
function of radial distance in tracing large-scale galaxy structures.

 
\subsection{Redshift distribution in the magnitude limit sample} 

By studying the redshift distribution in the Durham/UKST Galaxy
Redshift Survey, fluctuations have been found in the observed radial
density function are close to $50 \%$ occurring on $\sim 50$ Mpc/h
scales \citep{Ratcliffe98,busswell03}.  In a similar way in the 2dFGRS
\citep{busswell03}, two clear ``holes'' in the galaxy distribution
were detected in the ranges $0.03<z<0.055$, with an under-density of
$\sim 40 \%$, and $0.06<z<0.1$ where the density deficiency is $\sim
25 \%$.  These two under-densities, detected in particular in the the
2dFGRS southern galactic cap (SGC), are also clear features in the
Durham/UKST survey.  Given that the 2dFGRS SGC field is entirely
contained within the areas of sky observed for the Durham/UKST survey,
the similarities in the redshift distributions are both proofs of the
same features in the galaxy distribution \citep{busswell03}.

We can now compare the redshift distribution in the magnitude limit
sample with the results obtained by the SL analysis.  In
Fig.\ref{fig_counts_ml_ngp} we report the counting of galaxies as a
function of the radial distance, in bins of thickness 10 Mpc/h, in the
whole magnitude limited samples.  It is interesting to compare these
behaviors with Figs.\ref{fig_ngp400_sl}-\ref{fig_sgp550_sl}.  The SL
method clearly identifies the same structures, which are visible in
Fig.\ref{fig_counts_ml_ngp} as peaks of the radial
distribution. However the SL method is able to quantify the amplitude
of these fluctuations and, by applying the statistical analysis
presented above, to determine how typical these structures are.
\begin{figure}
\begin{center}
\includegraphics*[angle=0, width=0.5\textwidth]{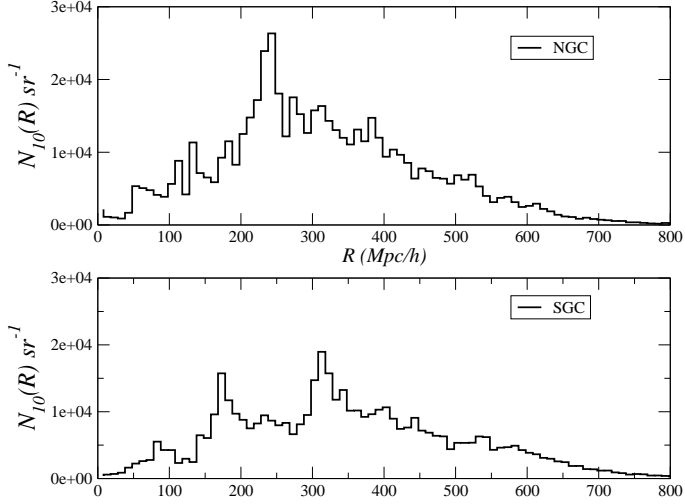}
\end{center}
\caption{{\it Upper panel}: Radial density in bins of thickness 10
  Mpc/h in the NGC magnitude limited sample.  The most prominent
  features identified by the $N(r;R)$ analysis are also visible by the
  simple counting. There is a large structure at $\sim 240$ Mpc/h.  By
  comparing this figure with
  Figs.\ref{fig_ngp400_sl}-\ref{fig_sgp550_sl} one may appreciate the
  usefulness of the SL method in tracing structures {\it Bottom
    panel}: The same for the SGC magnitude limited sample.  Here there
  are large structures at $\sim 180$ Mpc/h and $\sim 320$ Mpc/h.  }
\label{fig_counts_ml_ngp}
\end{figure}
%

 
\subsection{Magnitude counts}

In Fig.\ref{fig_counts_mag} we report the differential counts of
galaxies as a function of apparent magnitude in the SGC and NGC.  The
behaviors are similar to those found by \citet{norberg01,busswell03}:
the former paper concluded that there is  conclusive evidence that
counts in the SGC are down by $30\%$ relative to the NGC counts.  We
find the same difference and, as discussed, we can directly relate it
to the large-scale structures present in both
samples. Indeed, as already discussed, the amplitude of the
conditional number of galaxies (i.e., Eq.\ref{nr_est}) is $\sim 20
\div 30\%$ higher for the NGC samples than for the SGC ones.  In other
words, in the NGC samples there are more structures, hence 
fluctuations in the $N(r;R)$, than in SGC samples, as can be seen by
comparing, for instance, Figs.\ref{fig_ngp400_sl}-\ref{fig_sgp400_sl}.

\begin{figure}
\begin{center}
\includegraphics*[angle=0, width=0.5\textwidth]{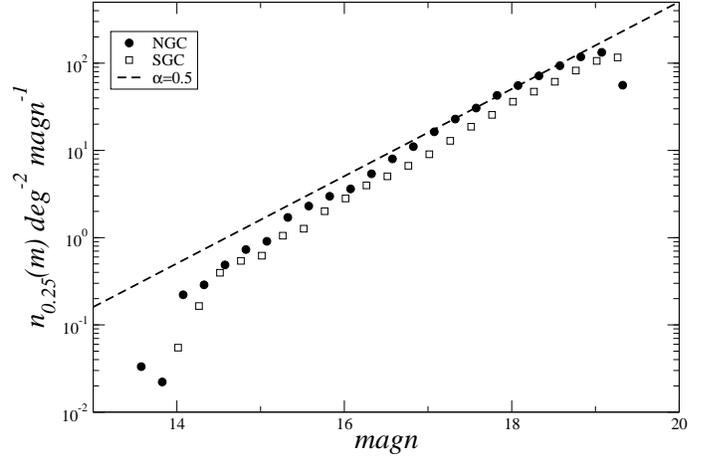}
\end{center}
\caption{Differential counts of galaxies, in bins of $\Delta m=0.25$,
  as a function of apparent magnitude in the SGC and NGC. A reference
  line corresponding to $N(m) \sim 10^{\alpha m}$ with $\alpha=0.5$ is
  reported.}
\label{fig_counts_mag}
\end{figure}


\subsection{The two-point correlation function}

The standard way to measure two-point correlations is accomplished by
determining the function $\xi(r)$ given by Eq.\ref{xi1} --- see e.g.,
\citet{tk69,dp83,park,benoist,
  zehavietal02,zehavietal04,norbergxi01,norbergxi02}.  As already
mentioned when measuring this quantity it is implicitly assumed that
the distribution is homogeneous well inside the sample volume;
i.e., $\lambda_0 \ll V^{1/3}$. Let us see what happens when, inside a
spherical sample of radius $R_s$ there is a fractal distribution with
dimension $D<3$.  We may estimate the sample density $n_S$ (which is
not an average quantity) by
\be
\label{aveden}
n_S = \frac{N}{V} = \frac{3B}{4\pi} R_s^{D-3}\;,
\ee
where, in the second equality on the r.h.s., we used Eq.\ref{nr_est}
for the number of points in spheres.  Equation \ref{aveden} shows that
the sample density depends on the sample size when $D<3$. The
estimator of the two-point correlation function can be written as
\citep{book}
\be
\label{xiestim} 
\overline{\xi(r)} +1 = \frac{\overline{N(r,\Delta r)}} {V(r,\Delta r)} \cdot
\frac{1} {n_S} \;.  
\ee 
The first ratio in the r.h.s. of Eq.\ref{xiestim} is the average
conditional density; i.e., the number of galaxies in shells of
thickness $\Delta r$ averaged over the whole sample, divided by the
volume $V(r,\Delta r)$ of the shell. The second ratio in the r.h.s. of
Eq.\ref{xiestim} is the sample density.  By using the second equality
on the r.h.s. of Eq.\ref{aveden} for $n_S$ and Eq.\ref{nr_est} for the
conditional number of points in spheres, we find
\be
\label{xiestfra} 
\overline{\xi(r)} = \frac{D}{3} \left( \frac{r}{R_s} \right)^{D-3} -1 \;,
\ee 
which shows that the amplitude of $\xi(r)$ depends on $R_s$ for $D<3$;
i.e., as long as the conditional density is a power law as a function
of scale \footnote{Pair-counting based estimators, like the Davis \&
  Peebles and/or the Landy \& Szalay estimator, have the same
  pathologies when the conditional density is a power law as a
  function of scale \citep{book}.} .

Equation~\ref{xiestfra} has been obtained by making the assumption
that the estimation of the sample density is given by the second
equality in the r.h.s. of Eq.\ref{aveden}. This is generallyot the
case, as the sample density in inhomogeneous distributions is
subjected to fluctuations of order one on the scale of the sample
size.  Thus the behavior given by Eq.\ref{xiestfra} should be
interpreted as giving a very rough estimation of the amplitude of
$\overline{\xi(r)}$ in a finite sample when there is a fractal
distribution inside it.

It is worth noticing that, if the conditional density is a power-law
function of scale, then $\overline{\xi(r)}$ {\it is not} a power-law
over the same of scales of the conditional density, and particularly
it does not have the same power law index . Indeed, as shown by
Eq.\ref{xiestfra}, $\overline{\xi(r)}$ has a a break of the power law
and it is possible to compute analytically the exponent of
$\overline{\xi(r)}$ as a function of the exponent of the conditional
density and the scale ratio $r/R_s$ \citep{book}.  It is easy to show
that on scales $r/Rs \ll 1$ the correlation exponent measured by the
two-point correlation analysis coincides with what is measured by the
conditional density, while on larger scales the exponent measured by
the $\xi(r)$ analysis is generally smaller than $D-3$. This is indeed
the result obtained by \citet{hawkins03} (see their Fig.6).

Let us now evaluate the sample density simply as $n_S=
N/V$ in SGC400 and NGC400.  Given that the sample geometry
is a sphere portion, the volume is given by 
\be 
V_{2dFGRS} = \frac{\Omega}{3} \left( R_{max}^3 - R_{min}^3 \right) \;.  
\ee 
By using the parameters of the VL samples (see
Table \ref{tbl_VLSamplesProperties}) we obtain respectively in the
SGC400 sample
\[
\overline{n_{sgc}} = 7.0 \cdot 10^{-3}
\] 
galaxies per (Mpc/h)$^3$ and in the NGC400 sample 
\[
\overline{n_{ngc}} = 1.0 \cdot 10^{-2} \;.
\] 
Thus there is a $\sim 30\%$ difference in the amplitude. When we
normalize the conditional density in the NGC400 and SGC400, which as
discussed have the same $\sim 30 \%$ difference in amplitude, to the
respective values of the sample density we get the amplitude of
$\overline{\xi(r)} $ is nearly the same (see Fig.\ref{xifs400}).  This
is because both the nominator and denominator in the r.h.s. of
Eq.\ref{xiestim} vary in the same way, having a difference of about
$30 \%$ in both samples.  Thus by measuring the amplitude of
fluctuations normalized to the sample density the information about
the large variations in the two different slices is lost. This {\it
  does not}, however, imply that fluctuations are small beyond $r_0
\approx 6$ Mpc/h as one would conclude by only considering the
two-point correlation function. Actually, as we discussed in the
previous sections, fluctuations are large and persistent over the
whole volume of these samples. It is indeed the normalization of
fluctuations amplitude to a not well-defined quantity, the sample
density, that in this case gives rise to the determination of the
length scale $r_0\approx 6$ Mpc/h, which is however a spurious
quantity.  The volume of the SGC400 sample is larger than that of the
NGC400 sample: this is reflected in the fact that the break of
$\overline{\xi(r)} $ occurs at a longer distance than for the NGC400
sample.
For the samples NGC550 and SGC550, there is a similar situation, but
the variation of both the sample density and of the amplitude of the
conditional density is of about $20\%$.
\begin{figure}
\begin{center}
\includegraphics*[angle=0, width=0.5\textwidth]{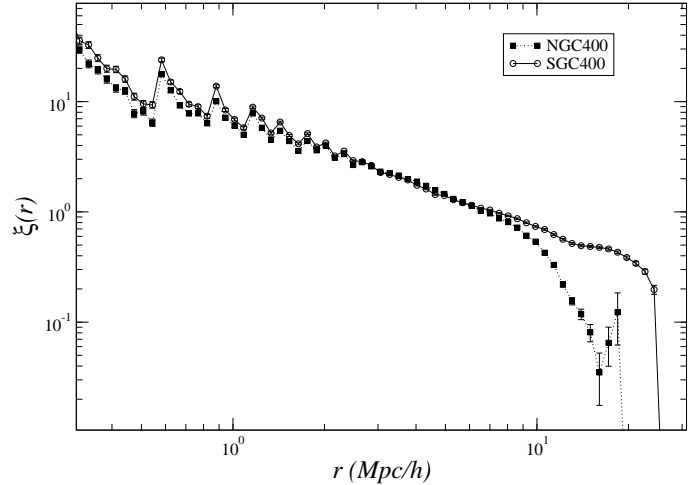}
\end{center}
\caption{Standard two-point correlation function in the SGC400 and
  NGC400 sample estimated by Eq.\ref{xiestim}. The sample density is
  simply computed as $N/V$.}
\label{xifs400}
\end{figure}

Different estimators of the two-point correlation function, such as
the Davis and Peebles (DP) \citep{dp83} estimator and the Landy and
Szalay (LS) \citep{landy93} estimator (see discussion in
\citet{cdm_theo} for more details about the different estimators),
lead to an estimation of the amplitude of $\overline{\xi(r)}$, which
agrees with the one just discussed above.  For instance in
Fig.\ref{xisgc400} a comparison is shown of the different
estimators. It is worth noticing that the different estimators give
the sample amplitude of $\overline{\xi(r)}$,  but they differ in the
scale at which $\overline{\xi(r)} $ has the break in the power-law
behavior. This is explained by the different ways the estimators treat
the boundary conditions, and, particularly, include (implicitly) the
global condition known as integral constraint --- see discussion in
\citet{cdm_theo}.
\begin{figure}
\begin{center}
\includegraphics*[angle=0, width=0.5\textwidth]{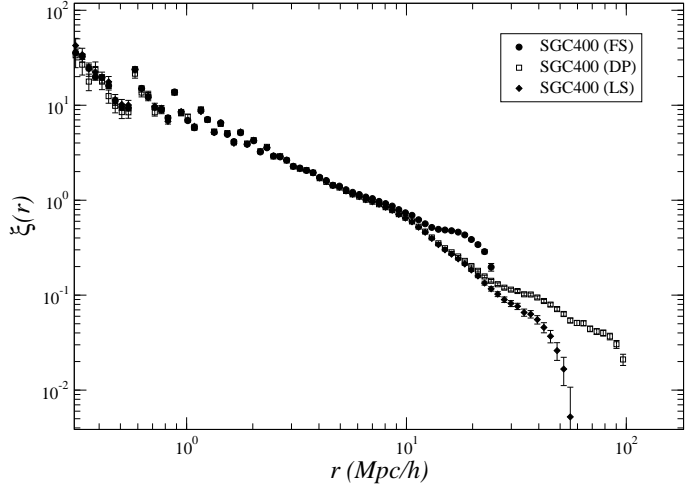} 
\end{center}
\caption{Standard two-point correlation function in the SGC400 measured 
by means of different estimators, namely the full-shell (FS),
the Davis and Peebles (DP)
and the Landy and Szalay (LS). } 
\label{xisgc400}
\end{figure}

The sample density $n_S$ can be estimated differently from
Eq.\ref{aveden}. The only condition that it is required for
Eq.\ref{xiestim} to be a valid estimator is that the size $r^*$ of the
volume entering in the denominator of Eq.\ref{aveden} be larger
than the homogeneity scale $\lambda_0$.  In fact,  when $r^* \gg
\lambda_0$ the estimation of the sample density does not differ
substantially from its ensemble average value because, in this
situation, the amplitude of the two-point correlation is, by
definition, much smaller than unity.  Thus we may consider another
estimator of the two-point correlation function, which is the one
introduced by \citet{paper1}
\be
\label{xi2est}
\overline{\xi (r;R,\Delta R)} +1 = \frac{\overline{N(r,\Delta r)}} {V(r, \Delta
  r)} \cdot \frac{V(R,\Delta R)}{\overline{N(R,\Delta R)}} \,, 
  \ee 
where the second ratio on the r.h.s. is the density of points in a
shell of thickness $r^* = \Delta R$ around the radial distance $R$.
If the distribution is homogeneous; i.e., $r^*>\lambda_0$, and
statistically stationary, Eq.\ref{xi2est} should be statistically
independent of the range of radial distances $(R,\Delta R)$
considered.  In Fig.\ref{xi_ngp500_sgp550} we show the determination
of Eq.\ref{xi2est} for different values of $R,\Delta R$ in the two
deepest samples. The amplitude of $\overline{\xi(r)}$ systematically
depends on the choice of the normalization and this is just the
imprint of the large scale inhomogeneities present in these samples.

In summary, even though the fact that there is a relatively large
difference in the densities between the NGC and the SGC, the amplitude
of the correlation function is similar because it is measured with
respect to a varying density; i.e., its value reflects the assumption
of homogeneity which is used in the definition in the $\xi$-analysis.
Only by analyzing fluctuations that are not normalized to the sample
density one can detect the effect of the large spatial inhomogeneities
characterizing these galaxy samples.

\begin{figure}
\begin{center}
\includegraphics*[angle=0, width=0.5\textwidth]{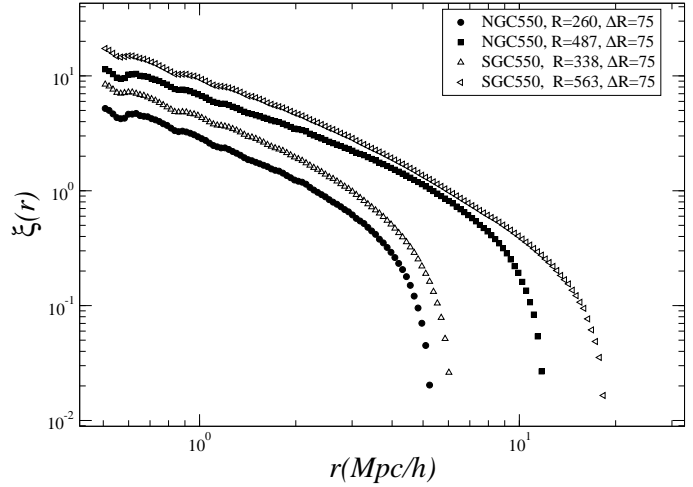}
\end{center}
\caption{Standard two-point correlation function in the SGC550 and
  NGC550 samples estimated by Eq.\ref{xi2est}. The sample average
  density is computed in spheres of radius $r^*$ and considering all
  center points lying in a bin of thickness $\Delta R$ centered at
  different radial distances $R$.}
\label{xi_ngp500_sgp550}
\end{figure}

\subsection{Analysis in the catalog with conservative cuts 
in apparent magnitude} 

We now briefly discuss the results of the SL analysis for the VL
samples obtained with conservative magnitude cuts (see
Table~\ref{tbl_VLSamplesProperties2}).  In Fig.\ref{fig1cons1} we show
the results of the SL analysis: by comparing this figure with
Figs.\ref{fig_ngp400_sl}-\ref{fig_sgp550_sl} by a simple visual
inspection one note that structures are extremely similar. Clearly,
because of the fewer points contained in the conservative-cuts
samples, the value of $N(r;R)$ is different. The behavior of the PDF
for the various samples is shown in Fig.\ref{fig1cons1}. The results
are thus statistically stable.
\begin{figure}
\begin{center}
\includegraphics*[angle=0, width=0.5\textwidth]{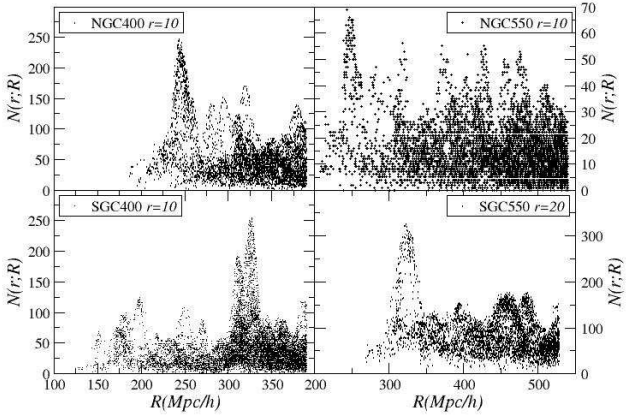}
\end{center}
\caption{SL analysis for the samples with conservative apparent
    magnitude cuts.  The value of the sphere radius is reported in the
    captions} 
\label{fig1cons1}
\end{figure}
\begin{figure}
\begin{center}
\includegraphics*[angle=0, width=0.5\textwidth]{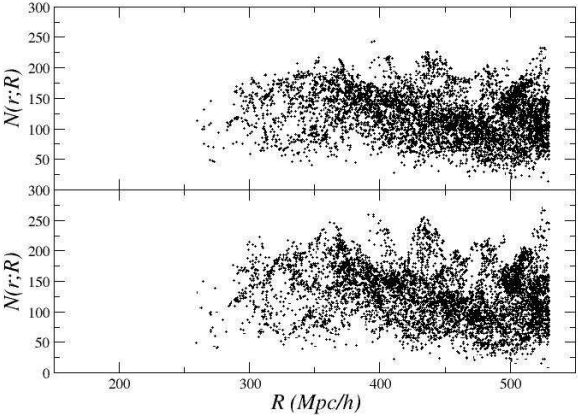}
\end{center}
\caption{PDF of the samples with conservative apparent magnitude cuts.
The value of the sphere radius is reported in the labels. In the comparison
we used the normalized variable $z$, by using the transformation 
described by Eqs.\ref{pdf_norm1}-\ref{pdf_norm2}}
\label{fig1cons2}
\end{figure}
%


\section{Comparison with mock-catalogs}
\label{sec:mock}

Standard theories of galaxy formation assume that fluctuations in the
matter density field in the early universe have very small amplitude,
\citep{pee80,peacock}. In this situation, by denoting as $P(k,t)$, the
PS of matter at an arbitrary time $t \gg 0$, where $t=t^*$ corresponds
to some early times in the universe as we discuss below, the
prediction of standard cosmological models can be written in this form
\be
\label{eqt3} 
P(k,t) = A P(k,t^*) g(k,t) \;.
\ee
It is possible to write $P(k,t^*) = P_{hz}(k) T^2(k)$ where
$P_{hz}(k)$ is the Harrison-Zeldovich (HZ) PS, and $T(k)$ is a
transfer function that depends on the type of coupling between matter
and radiation in the early universe \citep{peacock}.  The HZ PS is
predicted by inflationary theories to be the outcome of the
exponential amplification of quantum fluctuations occurring in the
early phases after the Big-Bang, and this is such that $P_{hz}(k)
\propto k $. A PS behaving as $P(k) \propto k^n$ on small $k$ with $n
\ge 1$ belonging to the class of super-homogeneous, or hyper-uniform
\citep{torquato}, distributions \citep{glass}. Their main
characteristic is that density fluctuations are of surface type; i.e.,
that the relative mass variance shows the fastest possible decay as a
function of scale $\sigma_{hz}^2(r) \propto 1/r^4$.  The PS must have
a cut-off at some high values of $k$; i.e, on small scales. This is
provided by the transfer function $T (k)$,  which can be determined
given the supposed properties of (dark) matter.  For CDM this is such
that at small $k$ the PS maintains the HZ tail, while at large $k$ it
 approximately decays as $k^{-2}$ \citep{peacock}. Correspondingly,
the two-point correlation function is positive for $r<r_c$ and
negative for $r>r_c$ with a power-law tail of the type $\xi(r) \sim -
r^{-4}$ \citep{cdm_theo}.

Gravitational clustering in the linear regime, in an Einstein de
Sitter cosmology, is characterized by a growing and a decaying mode,
both of them power laws in time \citep{pee80} $g(k,t) = g(t) \propto
t^{4/3}$. The amplitude $A$ in Eq.\ref{eqt3} is determined from the
observations of CMBR anisotropies and from the theoretical assumptions
on the nature of cosmological dark matter.  From the time dependence
of $P(k,t)$, it is possible to derive the time dependence of
$\lambda^m_0(t)$, defined to be $\sigma^2(\lambda^m_0)=1$. This
grows as a power-law function of time as well: particularly for
power-law PS; i.e., $P(k) \sim k^n$ and $n<4$, one obtains
\citep{pee80}
\be 
\lambda_0^m(t) \propto t^{\frac{3}{3(3+n)}} \;.  
\ee
To summarize the situation, we may identify two length-scales at the
present time. The first is the homogeneity (or nonlinearity) scale
$\lambda^m_0$: for $r<\lambda^m_0$ nonlinear clustering took place and
thus it changes the shape of PS in a non-linear way. On the other hand
for $r>\lambda^m_0$, fluctuations are still in the regime today, and
linear perturbation theory predicts a simple linear amplification of
primordial correlations.  From the normalization of the initial
amplitude $A$ to the CMBR anisotropies, $\lambda_0^m \approx 10$ Mpc/h
\citep{springel05,spergel}.  In addition, from the properties of the PS
of CMBR anisotropies it is derived that the second intrinsic length
scale of these models is about $r_c \approx 100$ Mpc/h
\citep{cdm_theo}.

In order to study gravitational structure formation in the nonlinear
phase, the common practice is to perform N-body simulations of
theoretical models.  This is done by integrating the equation of
motions of $N$ self-gravitating particles, in a volume $V$ and by
making use of periodic boundary conditions to represent an infinite
(periodic) system.  Initial particle correlations are given according
to a given theoretical model, and the initial redshift is generally
$z>10$. The simulation is then run up to $z=0$. In addition the space
background is expanding and thus one follows the particles' motion in
comoving coordinates. Particles are supposed to simulate the motion of
fluid elements of the underlying dark matter field.

To identify galaxies one uses a phenomenological approach. As
discussed, galaxies are supposed to form in the highest density peaks
of the dark matter field. Thus when the simulation has reached the
redshift $z=0$, one uses semi-analytic models to identify galaxies.
Among the largest simulations made publicy available, the millennium
run \citep{springel05} used more than 10 billion particles to trace
the evolution of the matter distribution in a region of the
universe in a cubic box of $500$ Mpc/h. Semi-analytic catalogs constructed
from the millennium run contains about 10 million objects
\citep{cronton06}.

Here we analyze the semi-analytic catalog containing 9,925,229 objects
in which the absolute magnitudes of mock-galaxies are given in the
BVRIK filters.  To reproduce the same limits in absolute magnitude of
the volume-limited samples of the 2dFGRS, we used the relations
between magnitude in different filters given by \citet{colless01}. In
this way we selected respectively 1,119,434 and 368,619 galaxies in a
500 Mpc/h cube.  We then selected three slices with the same geometry
of the real 2dFGRS samples, which are hereafter called SGC400m,
NGC400m and NGC550m.  The remaining sample, SGC550m, is $60^\circ$
wide instead of $84\circ$ as in real data. The number of objects in
each of the four samples is close to the one in the corresponding real
2dFGRS sample.

In Fig.\ref{SL_mock} we show the behavior of the SL analysis for the
mock-sample SGC550m in real and redshift space for $r=20$ Mpc/h (see
for comparison Figs.\ref{fig_ngp550_sl}-\ref{fig_sgp550_sl}).  The
effect of peculiar velocities is that of enhancing a little the
structures that appear in real space.  That the difference between
real and redshift space is small for sphere radius $r\ge 10$ Mpc/h is
shown in Figs.\ref{PNZ_400_10}-\ref{PNZ_550_20}, where we plot the PDF
of conditional fluctuations. In this case we  used the normalized
variable
\be
\label{pdf_norm1}
z_i(r) = \frac{N_i(r) - \overline{N(r)}}{\overline{\Sigma(r)}} \;,
\ee
and we thus determine its PDF, that is,
\be
\label{pdf_norm2}
P(z_i,r)=f\left(N_i(r)=\overline{N(r)}+z_i \overline{\Sigma(r)}\right)
\times \overline{\Sigma(r)} 
\ee
where $f(N,r)$ is the PDF of the variable $N_i(r)$, $\overline{N(r)}$
its estimated first moment, and $\overline{\Sigma(r)}$ its estimated
standard deviation on the scale $r$.
For sphere radius smaller than the homogeneity scale predicted by
theoretical models; i.e.,  $\lambda_0^m = 10$ Mpc/h, the PDF has a
relatively large tail for high values of $N$, although around the
peak it is well fitted by Gaussian function. For $r>10$ Mpc/h, the PDF
rapidly converges to a Gaussian and already for $r=20$ Mpc/h the fit
with a Gaussian function is extremely good also at high 
$N$. Correspondingly the plot of the SL analysis does not show the
presence of large amplitude fluctuations.

\begin{figure}
\begin{center}
\includegraphics*[angle=0, width=0.5\textwidth]{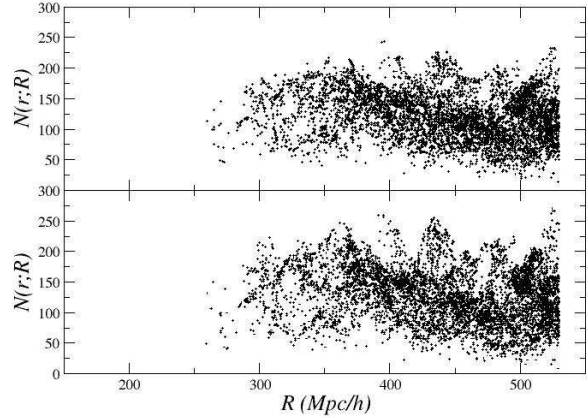}
\end{center}
\caption{SL analysis with sphere radius $r=20$ Mpc/, for the
  mock-sample SGC550m (the samples with better statistics) in real
  space ({\it upper panel}) and in redshift space ({\it bottom
    panel}). }
\label{SL_mock}
\end{figure}

\begin{figure}
\begin{center}
\includegraphics*[angle=0, width=0.5\textwidth]{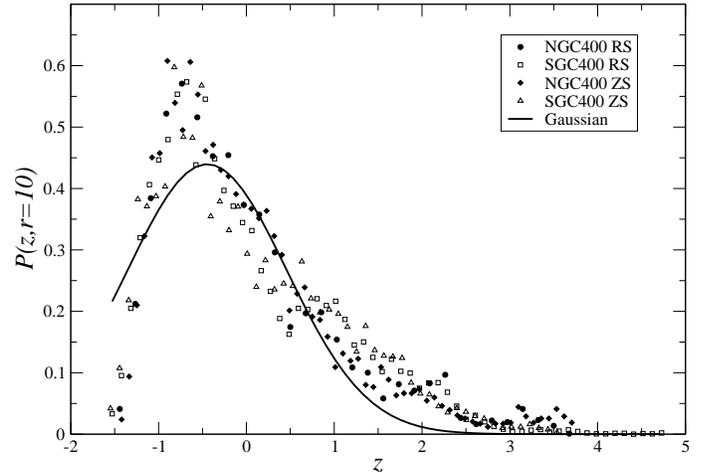} 
\end{center}
\caption{PDF for the mock-samples SGC400m and NGC400m in real space
    (RS) and redshift space (ZS) for $r=10$ Mpc/h. The best fit with a
    Gaussian function is reported as reference.}
\label{PNZ_400_10}
\end{figure}
\begin{figure}
\begin{center}
\includegraphics*[angle=0, width=0.5\textwidth]{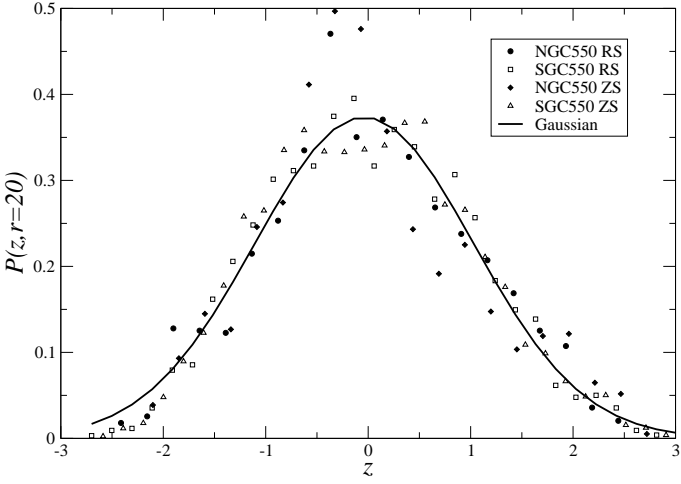} 
\end{center}
\caption{ The same as in Fig.\ref{PNZ_400_10} but for the case $r=20$
  Mpc/h and for the mock-samples SGC550m and NGC500m.}
\label{PNZ_550_20}
\end{figure}

In Figs.\ref{PNZ_10}-\ref{PNZ_20} we show the comparison between the
PDF in the mock-sample SGC500m (redshift space) and the one in the
real sample SGC550. Both for the sphere radius $r=10$ Mpc/h and $r=20$
Mpc/h, fluctuations are more persistent in real samples than in
mock-samples and this is reflected in the relatively fat tail of the
PDF for large $N$ values.
\begin{figure}
\begin{center}
\includegraphics*[angle=0, width=0.5\textwidth]{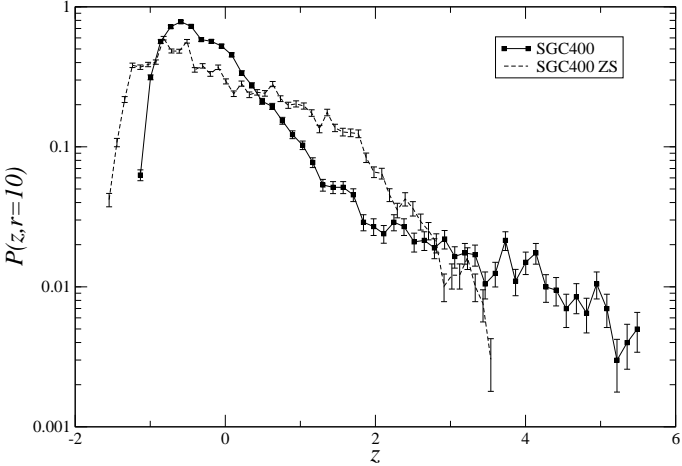} 
\end{center}
\caption{Comparison between the PDF in the mock-sample SGC400m
    (redshift space) and the one in the real sample SGC400. Poisson
    error bars are displayed as a reference.}  
\label{PNZ_10}
\end{figure}
\begin{figure}
\begin{center}
\includegraphics*[angle=0, width=0.5\textwidth]{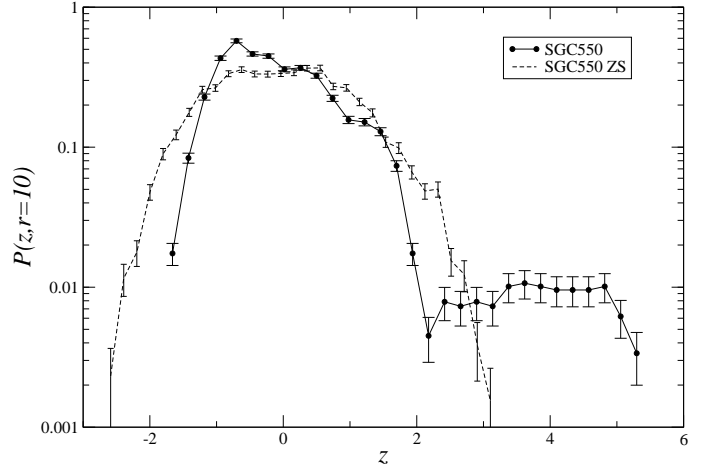} 
\end{center}
\caption{The same as in Fig.\ref{PNZ_10} but for the case $r=20$
    Mpc/h and for the sample SGC550.}  
\label{PNZ_20}
\end{figure}

In Fig.\ref{real_space_NR} (upper panel) it is shown the behavior of
the whole-sample average number of points is spheres;
i.e., Eq.\ref{eq1}. For $r > \lambda_0^m = 10$ Mpc/h, the function
$\overline{N(r)}$ growths with an exponent equal to the space
dimension (i.e., $D=3$). On smaller scales the exponent is instead
$D=1.2\pm 0.1$.

\begin{figure}
\begin{center}
\includegraphics*[angle=0, width=0.5\textwidth]{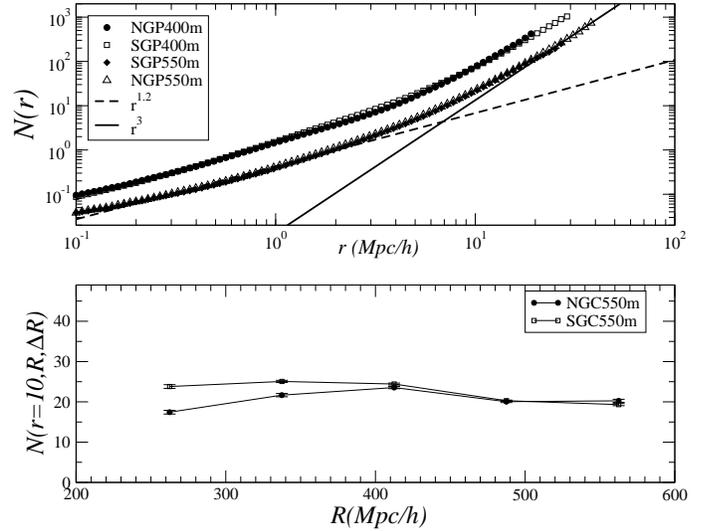}
\end{center}
\caption{{\it Upper panel}: Whole-sample average number of points in
  spheres; i.e., Eq.\ref{eq1} for the mock-samples in real space.  {\it
    Bottom panel}: Behavior of Eq.\ref{NRave} for the mock-galaxy
  samples in real space for SGC550m; the sphere radius is $r=10$ Mpc/h
  and the radial bin thickness is $\Delta R=75$ Mpc/h.  The error is
  computed through Eq.\ref{NRaveerr}.  }
\label{real_space_NR}
\end{figure}

Finally in the Fig.\ref{real_space_NR} (bottom panel) the behavior of
of Eq.\ref{NRave} with $r=5$ Mpc/h is shown for the mock-galaxy
samples in real space. Fluctuations, when averaged over radial bin of
thickness $\Delta R=50$ Mpc/h are more smooth and quiet than the real
samples and thus that there are no large fluctuations on scales larger
than $\lambda_0^m$. This is clearly consistent with the behavior of
the PDF described previously. The small difference for small $R$ is
due to the lower statistics at the nearby boundary of the samples.

Because the distribution becomes uniform at scales $r>\lambda_0\approx
10$ Mpc/h, the two-point correlation function is well-defined, both in
real and redshift space. This presents an intrinsic length scale
$\xi(r_0)=1$ at about $r_0 \approx 6$ Mpc/h, and beyond that scale it
rapidly decays to zero.

The whole-sample average number of points in spheres
(i.e., Eq.\ref{eq1}) does show a systematic difference with respect to
the real-space case. Indeed, while for $r>\lambda_0^m=10$ Mpc/h it
again shows a power-law behavior with $D=3$, on smaller scales the
exponent is $D=2.2\pm 0.1$ instead of $D=1.2$ as for the real space
case (see Fig.\ref{redshift_space_NR} --- upper panel).  In addition,
the transition between the two regimes is faster than for the real
space case.
\begin{figure}
\begin{center}
\includegraphics*[angle=0, width=0.5\textwidth]{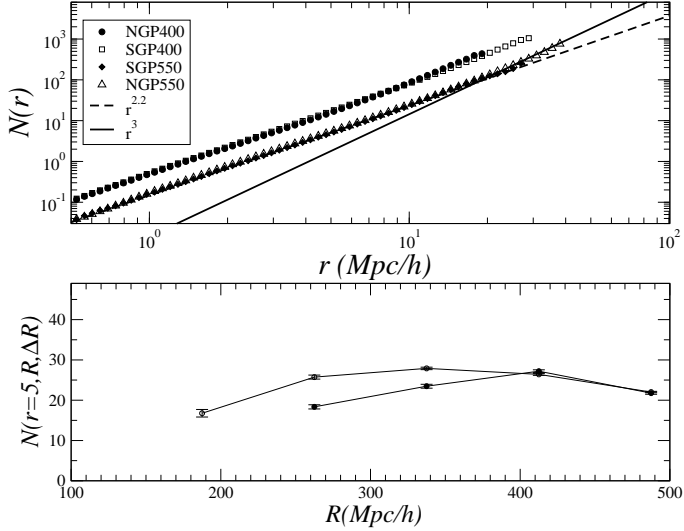}
\end{center}
\caption{{\it Upper panel}: Whole-sample average number of points in
  spheres; i.e., Eq.\ref{eq1}.  for the mock-samples in redshift space.
  {\it Bottom panel}: Behavior of Eq.\ref{NRave} for the mock-samples
  in redshift space for SGC550m; the sphere radius is $r=10$ Mpc/h and
  the radial bin thickness is $\Delta R=75$ Mpc/h.  The error is
  computed through Eq.\ref{NRaveerr}.  }
\label{redshift_space_NR}
\end{figure}

Finally in Fig.\ref{redshift_space_NR} (bottom panel) the behavior of
Eq.\ref{NRave} is reported for the mock-samples in redshift space.  As
for the real-space case, and consistent with the behaviors of the PDF
in redshift space, fluctuations are substantially smoothed when
averaged over radial bins of size $\Delta R > \lambda_0^m$.


\section{Discussion and conclusions}
\label{sec:discussion}

We characterized fluctuations and correlations in the galaxy
distribution, by considering several samples from the 2dFGRS
\citep{colless01}. The main point of our analysis is to use
statistical quantities that measure fluctuations amplitude not
normalizing it to the sample average. Our main results are: 

\begin{itemize}

\item Estimations of the sample average are affected by systematic
  variations that are the imprint of the structures with the large
  spatial extension present in these samples, which correspond to
  large amplitude fluctuations.  To identify structures in redshift
  space we applied the scale-length (SL) method, introduced in
  \citet{paper1},

\item The probability density function (PDF) of conditional
  fluctuations is, in all considered samples and up to $r=20$ Mpc/h
  (the largest distance scale we considered for this statistics),
  wildly different from a Gaussian function and characterized by a
  long tail, possibly with a power-law decay, signaling the presence
  of large fluctuations (e.g. structures) in the density field.  These
  fluctuations are persistent up to scales limited by sample sizes.

\item The first moment of the PDF shows scaling properties, that
  correspond to a fractal dimension $D=2.2 \pm 0.1$ up to $r \approx
  40$ Mpc/h. At the largest distance scale we considered; i.e., for
  $r>20$ Mpc/h, determinations are affected by relatively large
  fluctuations that reflect the intrinsic systematic error induced by
  structures, and thus the results have weaker statistical
  significance for the limited volume and number of points
  contributing to the average. This agrees with our previous findings
  in \citet{2df_paper}.

\item The homogeneity scale, i.e. the scale beyond which fluctuations
  become small and the sample density estimation approaches the value
  of the (ensemble) average density, has a lower limit at $\lambda_0 =
  70$ Mpc/h.  This is because fluctuations on scales of about 70 Mpc/h
  are incompatible with a transition to homogeneity on that scale.
  This result is consistent with the results by
  \citet{busswell03,Ratcliffe98}.  The main results obtained in this
  survey are in very good agreement with a similar analysis of galaxy
  samples extracted from the SDSS \citep{paper1}.

\item The SL method is suitable and effective in tracing structures in
  redshift space and in measuring the amplitude and distribution of
  density fluctuations.  Galaxy counts results can be easily related
  to the results obtained by the SL method. Indeed we confirmed that
  the variation of about $\sim 30\%$ between galaxy counts as a
  function of apparent magnitude in the northern and southern angular
  regions of this survey, detected by \citet{busswell03}, can be
  easily related to the different structures present in the different
  samples. Moreover, such large fluctuations, make any estimation of
  the sample average, on comparable scales to the samples sizes,
  affected by strong systematic effects.

 \item Apparently the amplitude the two-point correlation function,
   which normalizes the amplitude of fluctuations (as a function of
   scale) to the estimation of the sample average, of is not affected
   by fluctuations, and that this is stable in the Northern and
   Southern samples. This occurs, however, because of the
   normalization to the sample average (i.e., the denominator of
   $\xi(r)$) which indeed shows the same large variation between the
   two angular regions as the conditional density (i.e., the nominator
   of $\xi(r)$). Thus the normalization of fluctuations amplitude to
   the sample average hides the presence of large differences between
   the two samples, being biased by uncontrolled systematic
   finite-size effects. In this situation the amplitude of the
   two-point correlation function does not give a reliable and
   meaningful estimation of either the statistical properties of
   galaxy distribution in these samples, the spatial extension of
   correlations, or of amplitude of fluctuations.  It is indeed the
   normalization of fluctuations amplitude to a not well defined
   quantity, the sample density, which in this case determines the
   spurious length scale $r_0\approx 6$ Mpc/h.

\item We compared the results by studying conditional fluctuations in
  the 2dFGRS samples with those obtained in mock-galaxy
  catalogs. These are constructed from the dark matter density field
  in cosmological simulations which were evolved to study the effect
  of non-linear gravitational clustering in an expanding universe. We
  used the mock-galaxy samples provided by \citet{cronton06} and
  obtained from the largest N-body simulations publicly available at
  the moment \citep{springel05}.  We have studied the mock-galaxy
  catalogs both in real and redshift space, so to take into account
  the distortions due to the effect of peculiar velocities.  We found
  that the PDF of conditional fluctuations is weakly affected by
  peculiar velocities and that it rapidly converges to a Gaussian
  function on scales $r>10$ Mpc/h, in agreement with predictions of
  theoretical models both in real and redshift space. On smaller
  scales, i.e. $r<10$ Mpc/h, the PDF presents a longer tail than the
  Gaussian function. This is the signature of the nonlinear structures
  formed at small scales. Indeed, the PDF of mock-galaxy catalogs
  reflect the dynamics that has given rise to structures in the
  simulations.  In this respect, correlations provided by theoretical
  models and given as initial conditions for the simulations are of
  CDM type. In these models gravitational clutering forms nonlinear
  structures (i.e., large fluctuations in the density field) up to the
  limited range of about 10 Mpc/h. Such a length scale can be easily
  predicted by considering the initial amplitude of fluctuations in
  these models, normalized to CMBR anisotropies, and by computing the
  evolution in an expanding universe of the linear regime of
  gravitational clustering. Note that \citet{einasto} reached a
  similar conclusion; i.e., that in the 2dFGRS there are several rich
  super-clusters which is higher than what is found in similar volumes
  in the Millennium simulation.

\end{itemize}

Our conclusion is therefore that nonlinear structures (i.e., large
fluctuations) predicted by models extend to scales that are much
smaller than the ones we found in the 2dFGRS sample and that amplitude
of fluctuations on large scales, i.e., $r>10$ Mpc/h, is large in the
real samples and small in the mock-galaxy catalogs. Thus standard
models of galaxy formation, with CMBR normalization, are unable to
form large-scale structures such as the ones present in the observed
galaxy distribution.  In addition, standard models, beyond the value
of the homogeneity scale, predict a second intrinsic length scale
represented by the distance scale at which the two-point correlation
function changes sign \citep{bias,cdm_theo}; i.e., marking the
transition from small amplitude fluctuations with weak positive
correlations to the regime where there are anti-correlations (and
fluctuations of extremely small amplitude). This length scale is about
$r_c \approx 100$ Mpc/h in current models; i.e., slightly larger than
the lower limit to the homogeneity scale $\lambda_0 = 70$ Mpc/h we can
place by studying the 2dFGRS samples.  Forthcoming galaxy samples will
allow us to determine whether there will be a transition to
super-homogeneity (negative correlations) and on which scale this
eventually occurs.


\section*{Acknowledgments} We  thank Michael Joyce,
Andrea Gabrielli, and Luciano Pietronero for useful remarks and
discussions. We are grateful to Martin Lopez-Correidoira and Michael
Blanton for interesting comments. We also thank an anonymous referee
for useful suggestions. We acknowledge the use of the 2dFGRS data
available at {\tt http://www.mso.anu.edu.au/2dFGRS/ } and of the
millennium run semi-analytic galaxy catalog available at {\tt
  http://www.mpa-garching.mpg.de/galform/agnpaper/}.

{}


\begin{thebibliography}{aa}
 
 
\bibitem[Baryshev and Teerikorpi, 2005]{bt05} Baryshev, Yu.,
  Teerikorpi P., 2006, Bull.Spec.Astrop.Obs., 59, 92

\bibitem[Benoist et al., 1996] {benoist} Benoist, C., Maurogordato,
  S., da Costa, L. N., Cappi, A., Schaeffer, R., 1996, ApJ, 472, 452

 \bibitem[Busswell et al., 2004] {busswell03} Busswell, G.S. , et al.,
2004, MNRAS, 354, 991

\bibitem[Chiaki et al., 2003] {chiaki} Chiaki, H., et al., 2003, PASJ,
  55, 911

\bibitem[Colless et al., 2001]{colless01}
 Colless M., et al., 2001, MNARS. 328, 1039

\bibitem[Croton et al., 2006]{cronton06} Croton, D.J., et al., 
2006, MNRAS, 365, 11

\bibitem[da Costa et al., 1988]{ssrs2} da Costa, L. et al., 1988, ApJ,
  327, 544

\bibitem[Davis and Peebles, 1983] {dp83} Davis, M., Peebles, P.J.E.,
  1983, ApJ, 267, 465

\bibitem[Davis \& Huchra, 1982]{davishuchra} Davis, M. \& Huchra,
  J. 1982, ApJ, 254, 437


\bibitem[Davis et al., 1988]{davis88} Davis, M. et al., 1988, ApJ.,
  333, L9

\bibitem[de Lapparent Huchra and Geller, 1986]{dlhg85} 
de Lapparent, V., Geller, M. J., Huchra, J. P., 
1986, ApJ, 302, 1


\bibitem[de Vaucouleurs, 1970]{devac1970}
de Vaucouleurs, G., 1970, Science, 167, 1203

\bibitem[Durrer et al., 2003]{bias} Durrer, R., Gabrielli, A., Joyce,
  M., Sylos Labini, F., 2003, ApJ, 585, L1

\bibitem[Eke et al., 2004]{eke04} Eke, V., et al., 2004 MNRAS, 348, 866

\bibitem[Einasto et al., 2006]{einasto}
Einasto, J., et al., 2006, A\&A, 459, L1 

\bibitem[Falco et al., 1999]{cfa2} Falco, E.F., et al., 
1999, PASP, 111,  438, 


\bibitem[Frith et al., 2003]{frith03}  
Frith, W.J. ,
G.S. Busswell, R. Fong, N. Metcalfe, T. Shanks,
2003,  MNRAS,   345, 1049
 
\bibitem[Gabrielli and Sylos Labini, 2001]{gsl00} Gabrielli, A., Sylos
  Labini, F., 2001, Europhys.Lett., 54, 286

\bibitem[Gabrielli et al., 2002]{glass} Gabrielli, A., Joyce, M.,
   Sylos Labini, F., 2002, Phys.Rev., D65, 083523

\bibitem[Gabrielli et al., 2005]{book} Gabrielli A., Sylos Labini F.,
  Joyce M., Pietronero L., 2005 Statistical Physics for Cosmic
  Structures (Springer Verlag, Berlin)

\bibitem[Geller and Huchra, 1989]{gh89} Geller, M., Huchra J., 1989,
  Science 246, 897

\bibitem[Giovanelli and Haynes, 1993]{pp}
Giovanelli, R.,  Haynes, M. P.,1993, AJ, 105, 1271

\bibitem[Gott et al., 2005]{sloangreatwall} Gott, J.R. III, et al.,
  2005, Ap.J., 624, 463

\bibitem[Hawkins et al., 2003]{hawkins03} Hawkins E., et al., 2003,
MNARS, 346, 78

\bibitem[Hogg et al., 2005]{hogg}
Hogg D.W., Eisenstein D., Blanton M. et al., 2005, ApJ, 624, 54 

\bibitem[Hogg et al., 2002]{hogg_kterm} Hogg, D.W., Baldry, I.K.,
  Blanton, M.R., Eisenstein, D.J., 2002, eprint
  arXiv:astro-ph/0210394v1

\bibitem[Huchra et al., 1983]{cfa1} 
Huchra, Davis, Latham and Tonry, 1983, ApJS 52, 89

\bibitem[Kaiser, 1984]{kaiser} Kaiser, N., 1984, ApJ, 284, L9

\bibitem[Kerscher et al., 1999] {kerscher98} Kerscher, M., Schmalzing,
  J., Buchert, T. and Wagner, H., 1998 A\&A, 333, 1

\bibitem[Kerscher, 1999]{kerscher} 
Kerscher, M. 1999, A\&A, 343, 333 

\bibitem[Joyce and Sylos Labini, 2001] {jsl} Joyce, M., Sylos Labini,
  F., 2001, ApJ, 554, L1

\bibitem[Landy and Szalay, 1993]{landy93}
Landy S. D., Szalay A. 1993, ApJ, 412, 64

\bibitem[Madgwick et al., 2002]{madgwick02} Madgwick D. S., et
  al. 2002, MNRAS, 333, 133

\bibitem[Norberg et al., 2002a]{norberg01} Norberg, P., et al., 2002,
  MNRAS, 306, 907

\bibitem[Norberg et al., 2001]{norbergxi01} Norberg E.,  et al.,
2001, MNRAS, 328, 64

\bibitem[Norberg et al., 2002b]{norbergxi02} Norberg E., et al., 2002,
  MNRAS, 332, 827

\bibitem[Park et al., 1994]{park} Park, C., Vogeley, M. S., Geller,
  M. J., Huchra, J. P., 1994, ApJ, 431, 569

\bibitem[Peacock, 1999]{peacock} Peacock, J.A., 1999, Cosmological
  physics (Cambridge University Press)

\bibitem[Peebles, 1980] {pee80}
Peebles P. J. E., 1980, The Large-Scale Structure of the
Universe, (Princeton University Press)

\bibitem[Picard, 1991]{picard91} Picard, A.,1991, AJ, 102, 445

\bibitem[Ratcliffe et al., 1998]{Ratcliffe98} Ratcliffe, A., Shanks,
  T., Parker, Q.A. \& Fong, R., 1998a, MNRAS, 293, 197

\bibitem[Saslaw, 2000]{saslaw} Saslaw, W.C., 2000, 
The Distribution of the Galaxies,
(Cambridge University Press,  Cambridge)

\bibitem[Shectman et al., 1996]{lcrs} Shectman, S. A., et al., 
1996, ApJ, 470, 172 

\bibitem[Spergel et al., 2007] {spergel} Spergel, D.N., et al., 2007,
  ApJS, 170, 377

\bibitem[Springel et al., 2005] {springel05} Springel, V., et al.,
  2005, Nature, 435, 629

\bibitem[Sylos Labini et al., 1998]{slmp98} Sylos Labini, F.,
  Montuori, M., Pietronero, L., 1988, Phys.Rep., 293, 61

\bibitem[Sylos Labini et al., 2007]{sdss_paper} 
Sylos Labini, F., Vasilyev, N.L.  Baryshev, Yu.V.,
2007, A\&A, 465, 23 

\bibitem[Sylos Labini et al., 2008]{paper1} Sylos Labini, F.,
  Vasilyev, N.L., Pietronero, L., Baryshev, Yu. V., 2008, eprint
  arXiv:0805.1132v1

\bibitem[Sylos Labini and Vasilyev, 2008] {cdm_theo} Sylos Labini, F.,
  Vasilyev, N.L., 2008, A\&A, 477, 381

\bibitem[Sylos Labini et al., 2009]{2df_2009} 
Sylos Labini, F., Vasilyev, N.L.  Baryshev, Yu.V.,
2009, Europhys.Lett., in the press

\bibitem[Torquato and Stillinger, 2003]{torquato} 
Torquato, S. \&  Stillinger, F. H., 2003, 
Phys. Rev., E, 68, 041113 
 
\bibitem[Totsuji and Kihara, 1969] {tk69} Totsuji, H., Kihara, T.,
  1969, PASJ, 21, 221

\bibitem[Vasilyev et al., 2006]{2df_paper} Vasilyev, N.L., Baryshev,
  Yu. V., Sylos Labini, F., 2006, A\&A, 447, 431

\bibitem[York et al., 2000]{sdss}  York, D., et al., 2000, AJ,  120, 1579 

\bibitem[Weir et al., 1995]{gd} 
Weir N., Djorgovski, G.S., Fayyad, U.M., 
1995, AJ, 110, 1

\bibitem[Wu et al., 1999]{rees} Wu, K.K., Lahav, O. \& Rees, M., 
1999, Nature,  225, 230

\bibitem[Zehavi et al., 2002]{zehavietal02} Zehavi, I., et al., 2002,
  ApJ, 571, 172

\bibitem[Zehavi et al., 2004]{zehavietal04} Zehavi, I., et al., 2004,
  ApJ, 608, 16


\bibitem[Zehavi et al., 2005]{zehavi05} 
Zehavi, I., et al., 2005, ApJ, 621, 22

\end{thebibliography}
\end{document}